%% file: main.tex
\documentclass[conference]{IEEEtran}

\ifCLASSINFOpdf
\else
\fi

\usepackage[normalem]{ulem}
\usepackage{wrapfig}
\usepackage{lipsum}

\usepackage[parfill]{parskip}
\usepackage{booktabs} 
\usepackage{float}
\usepackage{graphicx}
\usepackage{epstopdf}
\usepackage{epsfig}
\usepackage[font=small,subrefformat=parens]{subcaption}
\usepackage{lipsum}
\usepackage{multirow,multicol,array}
\usepackage{tabularx}
\usepackage{pifont}
\usepackage{setspace}
\usepackage{mwe}
\usepackage{algorithm,algpseudocode,amsmath}

\usepackage{array}
\newcolumntype{P}[1]{>{\centering\arraybackslash}p{#1}}
\newcolumntype{M}[1]{>{\centering\arraybackslash}m{#1}}
\DeclareCaptionLabelFormat{andtable}{#1~#2  \&  \tablename~\thetable}
\DeclareCaptionLabelFormat{andfigure}{\tablename~\thetable \ \&  #1~#2}

\algnewcommand\algorithmicforeach{\textbf{for each}}
\algdef{S}[FOR]{ForEach}[1]{\algorithmicforeach\ #1\ \algorithmicdo}

\newlength{\bibitemsep}
\newlength{\bibparskip}
\let\oldthebibliography\thebibliography
\renewcommand\thebibliography[1]{%
  \oldthebibliography{#1}%
  \setlength{\parskip}{\bibitemsep}%
  \setlength{\itemsep}{\bibparskip}%
}

\makeatletter
\def\blfootnote{\gdef\@thefnmark{}\@footnotetext}
\makeatother

\include{macros}

\makeatletter
\def\footnoterule{\kern-3\p@
  \hrule \@width 2in \kern 2.6\p@} 
\makeatother

\begin{document}
\title{Understanding System Characteristics of Online Erasure Coding on Scalable, Distributed and Large-Scale SSD Array Systems}

\author{\IEEEauthorblockN{Sungjoon Koh\IEEEauthorrefmark{1},
Jie Zhang\IEEEauthorrefmark{1}, Miryeong Kwon\IEEEauthorrefmark{1}, Jungyeon Yoon\IEEEauthorrefmark{3}, David Donofrio\IEEEauthorrefmark{2},\\
 Namsung Kim\IEEEauthorrefmark{4} and Myoungsoo Jung\IEEEauthorrefmark{1}}
\IEEEauthorblockA{\IEEEauthorrefmark{1}Computer Architecture and Memory Systems Laboratory,\\ School of Integrated Technology, Yonsei Institute Convergence Technology, Yonsei University
\\\IEEEauthorrefmark{2}Lawrence Berkeley National Laboratory,\IEEEauthorrefmark{3}SK Telecom,\IEEEauthorrefmark{4}University of Illinois Urbana-Champaign\\
\small\IEEEauthorrefmark{1}skoh@camelab.org,\IEEEauthorrefmark{1}jie@camelab.org,\IEEEauthorrefmark{1}mkwon@camelab.org,\IEEEauthorrefmark{3}jungyeon.yoon@sk.com,\IEEEauthorrefmark{2}ddonofrio@lbl.gov,
\\\IEEEauthorrefmark{4}nskim@illinois.edu,\IEEEauthorrefmark{1}mj@camelab.org
}\vspace{-25pt}}

\maketitle

\begin{abstract}
Large-scale systems with arrays of solid state disks (SSDs) have become increasingly common in many computing segments. 
To make such systems resilient, we can adopt erasure coding such as Reed-Solomon (RS) code as an alternative to replication 
because erasure coding can offer a significantly lower storage cost than replication.
To understand the impact of using erasure coding on system performance and other system aspects such as CPU utilization and network traffic,
we build a storage cluster consisting of approximately one hundred processor cores with more than fifty high-performance SSDs, 
and evaluate the cluster with a popular open-source distributed parallel file system, Ceph. 
Then we analyze behaviors of systems adopting erasure coding from the following five viewpoints, compared with those of systems using replication: 
(1) storage system I/O performance;
(2) computing and software overheads; 
(3) I/O amplification; 
(4) network traffic among storage nodes; 
(5) the impact of physical data layout on performance of RS-coded SSD arrays. 
For all these analyses, we examine two representative RS configurations, which are used by Google and Facebook file systems, and compare them with triple replication that a typical parallel file system employs as a default fault tolerance mechanism. 
Lastly, we collect 54 block-level traces from the cluster and make them available for other researchers.
\let\thefootnote\relax\footnote{This paper is accepted by and will be published at 2017 IEEE International Symposium on Workload Characterization. This document is presented to ensure timely dissemination of scholarly and technical work.}
\end{abstract}

\setstretch{0.985}

\vspace{-8pt}
\section{Introduction}
\vspace{-8pt}
\label{sec:intro}
\input{intro}

\vspace{-8pt}
\section{Background}
\label{sec:background}
\input{background}

\vspace{-10pt}
\section{Evaluation Methodology}
\label{sec:evalsetup}
\input{evalsetup}

\vspace{-8pt}
\section{Overall Performance Comparison}
\vspace{-3pt}
\label{sec:overall}
\input{overall}

\vspace{-8pt}
\section{Testing Computation and System}
\label{sec:system}
\vspace{-7pt}
\input{system}

\vspace{-5pt}
\section{Testing Data Overheads}
\label{sec:dataoverhead}
\vspace{-5pt}
\input{dataoverhead}

\vspace{-7pt}
\section{Testing Data Distribution}
\vspace{-5pt}
\label{sec:distribution}
\input{distribution}

\vspace{-11pt}
\section{Related Work and Discussion}
\vspace{-11pt}
\label{sec:discussion}
\input{discussion}

\vspace{-10pt}
\section{Acknowledgements} 
\vspace{-8pt}
\label{sec:acknowledgement}
\input{acknowledgement}

\vspace{-10pt}
\section{Conclusion} 
\vspace{-9pt}
\label{sec:conclusion}
\input{conclusion}


\end{document}

%% file: macros.tex
\newcommand{\mycomment}[1]{}
\newcommand{\ignore}[1]{}

\usepackage{color}
\usepackage{soul}
\usepackage{xspace}



%% file: intro.tex
Due to the explosion of data across all market segments, there are strong demands for scalable and high-performance distributed storage systems. 
Recently, high performance computing (HPC) and data center (DC) systems have begun to adopt distributed storage systems comprising powerful computing resources with arrays of many solid state drives (SSDs) instead of traditional spinning hard disk drives (HDDs). 
The latency and bandwidth of SSDs in such distributed storage systems are approximately 2$\times$ shorter and higher than those of enterprise HDDs, while SSDs consume much less power than HDDs. 
Such low-power and high-performance characteristics of SSDs is desirable for scalable and high-performance distributed storage systems.

SSD arrays in distributed storage systems can accelerate block I/O services and make many latency-sensitive applications better perform tasks processing a large amount of data.
However, existing configurations of distributed storage systems are not well optimized for SSDs. Especially, distributed storage systems require strong fault tolerance mechanisms since storage devices often suffer from failures. For example, Facebook reports that up to 3\% of its storage devices fail every day \cite{sathiamoorthy2013xoring}. 
Although the reliability of SSDs is much higher than that of HDDs, such frequent failures should be efficiently handled. 
In addition to the failures of disks themselves, hardware and software failures of network switches and stroage nodes caused by soft errors, hard errors and/or power outages also prevent accesses of some storage devices in storage nodes \cite{greenberg2008cost}.
To keep data available and protected against both such software and hardware failures, therefore conventional distributed storage systems often employ replication, which is a simple but effective method to make distributed storage systems resilient. However, replicating a large amount of data can incur significant storage overheads and performance degradations  \cite{hu2009write, jung2013revisiting, jung2016exploring}.

Erasure coding becomes a fault tolerance mechanism alternative to replication as it can offer the same reliability as or higher reliability than triple replication (denoted by ``3-replication''), with much lower storage overheads. 
The Reed-Solomon (RS) code is one of the most popular erasure codes due to its optimal storage space utilization \cite{reed1960polynomial, rashmi2015having, rashmi2016ec, mitra2016partial} and can be easily applied to SSD arrays in distributed storage systems to address the overheads imposed by traditional replication. 
When an RS code is employed, all input data are stored as \emph{chunks} with a fixed size. 
In general, RS($k$,$m$) computes $m$ code chunks for $k$ data chunks, and the system distributes the data and code chunks across different storage devices or nodes, referred to as \emph{stripe}. 
For chunks belonging to the same stripe, RS($k$,$m$) can repair failures in up to $m$ data chunks. 
For example, Google Colossus, which is the successor of the Google File System \cite{metz2012google, ghemawat2003google, ford2010availability}, uses RS(6,3) to tolerate any failure in up to three chunks with only 1.5$\times$ storage overheads. 
Compared with 3$\times$ storage overheads of traditional replication, 1.5$\times$ storage overheads of RS(6,3) is attractive for distributed storage systems.

\vspace{-3pt}
However, one of disadvantages in applying erasure coding to distributed storage systems is high pressures of their reconstruction operations on system performance. 
When a node discovers that a data chunk is missing due to a failure, RS($k$,$m$) requires the node to bring $k-1$ remaining chunks over the network, reconstruct the missing chunk, and send the repaired chunk to the corresponding node (i.e., $k\times$ more network traffic). 
Such a network overhead, also referred to as \emph{repair traffic}, is a well-known issue \cite{dimakis2011survey, sathiamoorthy2013xoring}. 
For example, a Facebook cluster deploying erasure coding increases network traffics by more than 100TB of data transfer in a day associated with data reconstruction \cite{rashmi2013solution}. 
To address this problem, there have been many studies on optimal trade-offs between network traffic and storage overheads for repairs \cite{huang2012erasure, sathiamoorthy2013xoring, esmaili2013core}. 
Exploiting such a trade-off between network traffic and storage overheads for repairs (e.g., lower storage overhead at the cost of higher repair network traffic)\cite{huang2012erasure, sathiamoorthy2013xoring}, system architects can choose an optimal coding scheme depending on a given system architecture.

\begin{figure}
	\begin{center}
		\includegraphics[width=\linewidth]{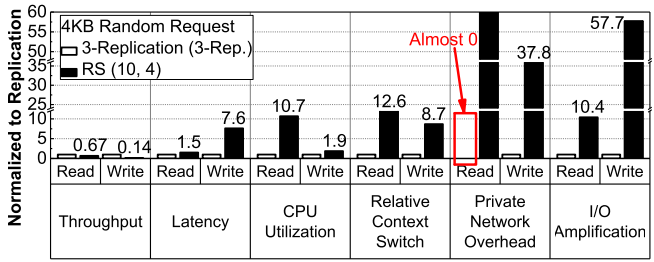}
	\end{center}
	\vspace{-6pt}
	\caption{Comparison of replication and erasure coding approaches; this figure summarized the performance and overhead analysis that this work conducted in terms of 6 different viewpoints \vspace{-15pt}} 
	\label{fig:introduction}
\end{figure}

\vspace{-3pt}
There have been many studies on the repair traffic, but a little attention has been paid to online usages of RS codes in distributed storage systems. 
The repair traffic occurs only when there is a failure, 
but encoding and decoding data always incur various overheads such as more CPU cycles, network traffic and I/O amplification in SSDs due to RS encoding and decoding.
Furthermore, since SSDs are much faster than HDDs, data need to be encoded/decoded and transferred over network at higher rates to reap the benefits of using SSDs.
This motivates us to study the overheads on the network and storage nodes in detail.
Such a study will allow us to more efficiently 
use distributed storage systems based on SSD arrays. 

\vspace{-3pt}
In this work, we build a distributed storage system consisting of 96 processor cores and 52 high-performance SSDs. 
We leverage an open-source parallel file system, Ceph \cite{weil2006ceph}, which is widely used for distributed storage systems, 
because it can easily support erasure coding as a plug-in module. 
We examine the distributed storage system by evaluating diverse I/O-intensive workloads with two popular RS configurations, RS(6,3) and RS(10,4) used by Google and Facebook file systems \cite{metz2012google, ford2010availability, borthakur2010hdfs}, 
and compare the behaviors of systems deploying these two RS configurations with those of a system adopting 3-replication. 
To the best of our knowledge, this work\footnote{All the traces collected for this paper are available for download from http://trace.camelab.org.} is the first examination of various overheads imposed by online erasure coding on a real-world distributed storage system with an array of SSDs. 
Our observations and findings through in-depth analyses can be summarized as follows:

\vspace{-3pt}
\noindent $\bullet$ \textbf{Storage system I/O performance overhead.} This work comprehensively analyzes the impacts of online RS coding on I/O performance of storage systems, and compares them with those of the default fault tolerance scheme (i.e., 3-replication). 
As shown in Figure \ref{fig:introduction}, even without any failure, RS(10,4) coding gives 33\% lower bandwidth and 50\% latency than 3-replication for read operations respectively. 
Furthermore, RS(10,4) coding exhibits
86\% lower bandwidth and 7.6$\times$ longer latency than 3-replication for write operations. 
We believe that this performance degradation imposed by erasure coding will be a significant challenge and make it difficult to deploy SSD arrays for HPC and DC systems. 

\vspace{-3pt}
\noindent $\bullet$ \textbf{Computing and software overheads.} 
In each node, erasure coding is mostly implemented with software as part of a RAID virtual device driver in kernel space or with hardware as part of RAID controllers \cite{ref-raid}. 
However, since most parallel file systems are implemented in user space, erasure coding for distributed storage systems often requires to perform I/O services by passing I/O requests through all the modules of the storage stack at both user and kernel spaces. 
Consequently, this user-space implementation of erasure coding can increase computing overheads, including context switching, encoding data through a generator matrix, and data placement. 
Figure \ref{fig:introduction} shows the comparison of overall CPU utilizations of 3-replication with those of RS(10,4) coding. 
We can observe that RS(10,4) coding consumes 10.7$\times$ more CPU cycles than 3-replication. 
In this work, we perform an in-depth study for computing and software overheads of online erasure coding by decomposing them into user- and kernel-level activities. 

\begin{figure*}
\centering
\includegraphics[width=1\linewidth]{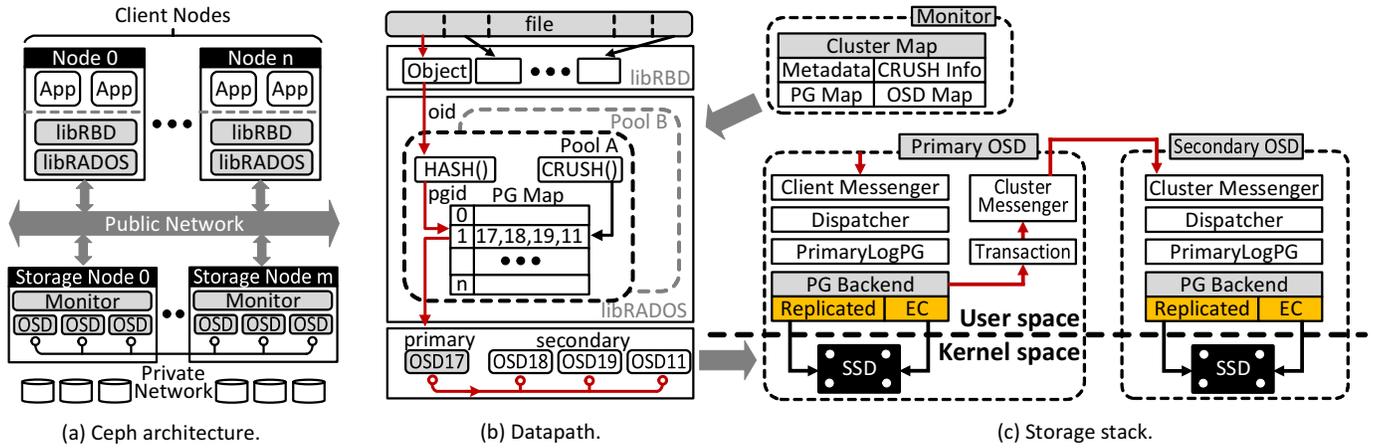}
\vspace{-20pt}
\caption{Overview of distributed storage systems with SSD arrays and a Ceph parallel file system. \vspace{-15pt}}
\label{fig:ceph}
\end{figure*}

\noindent $\bullet$ \textbf{I/O amplification overheads.} We observe that online erasure coding significantly increases the amount of I/O data, which are served by storage nodes, compared with 3-replication. 
This happens even without any failure. 
As shown in Figure \ref{fig:introduction}, RS(10,4) coding reads 10.4$\times$ more data from the underlying storage devices than 3-replication.
This is because RS coding manages all I/O requests at the stripe level, which is 
inefficient from the storage viewpoint. 
For instance, we also observe that RS(10,4) coding writes 57.7$\times$ more data to the underlying Flash media than 3-replication for random writes. 
In contrast to the common expectation for using erasure coding (e.g., significantly lower storage overheads), 
erasure coding needs to read/write a significantly larger amount of data from/to Flash media than 3-replication. 
Since Flash media does not allow an overwrite to a physical block (without erasing), SSD firmware forwards the incoming write to another reserved block, which was erased in advance, and invalidate out-of-data by remapping the corresponding address with new one \cite{zhang2015opennvm, shahidi2016exploring}.   
These Flash characteristics increase the actual amount of write data and thus shorten the life time of SSD when erasure coding is used, 
irrespective of the actual amount of data that erasure coding manages. 

\noindent $\bullet$ \textbf{Private network overheads.} 
In contrast to erasure coding implemented for a single node, 
erasure coding for distributed storage systems pulls or pushes a large amount of data over its private network which connects storage nodes and is invisible to client nodes. 
Figure \ref{fig:introduction} shows the amount of network traffic that erasure coding generates, normalized to that of network traffic that 3-replication generates. 
Due to the coding characteristics, RS coding generates a significant amount of read traffic for write operations in contrast to 3-replication. 
More importantly, erasure coding congests the private network due to stripes and data regions that erasure coding needs to manage for coding chunks, whereas 3-replication hardly congests the network. 
Even for reads, which requires no decoding, the private network of a system deploying erasure coding is much busier than that of a system using 3-replication because erasure coding needs to concatenate chunks. 
Such high private network traffic is necessary to recovery from errors in one or more storage nodes, but we believe that higher network overheads of erasure coding than replication should be optimized for future distributed storage systems while especially considering much shorter latency of SSDs than HDDs.

\noindent $\bullet$ \textbf{Physical data layout on RS-coded SSD array.} 
Since there are many SSD devices in a distributed storage system and all of them are connected by network, 
the performance of erasure coding also significant changes depending on the layout of the data across SSD devices of storage nodes. 
In practice, the performance of an SSD degrades if reads and writes are interleaved and/or there are many random accesses \cite{chen2009understanding}. 
In contrast to this device-level characteristic, we observe that the throughput of data updates in erasure coding (each of which consists of reads and writes due to the parity chunks) is 1.6$\times$ better than that of a new write (encoding). In addition, we also observe that, the performance of random accesses is better than that of sequential accesses due to the data layout of underlying SSD arrays. 
In this work, we will examine the system and performance characteristics of RS codes by considering the physical data layout that the parallel file system manages.

%% file: background.tex

%

\begin{figure*}
\centering
\begin{subfigure}[b]{0.23\linewidth}
\includegraphics[width=1\linewidth]{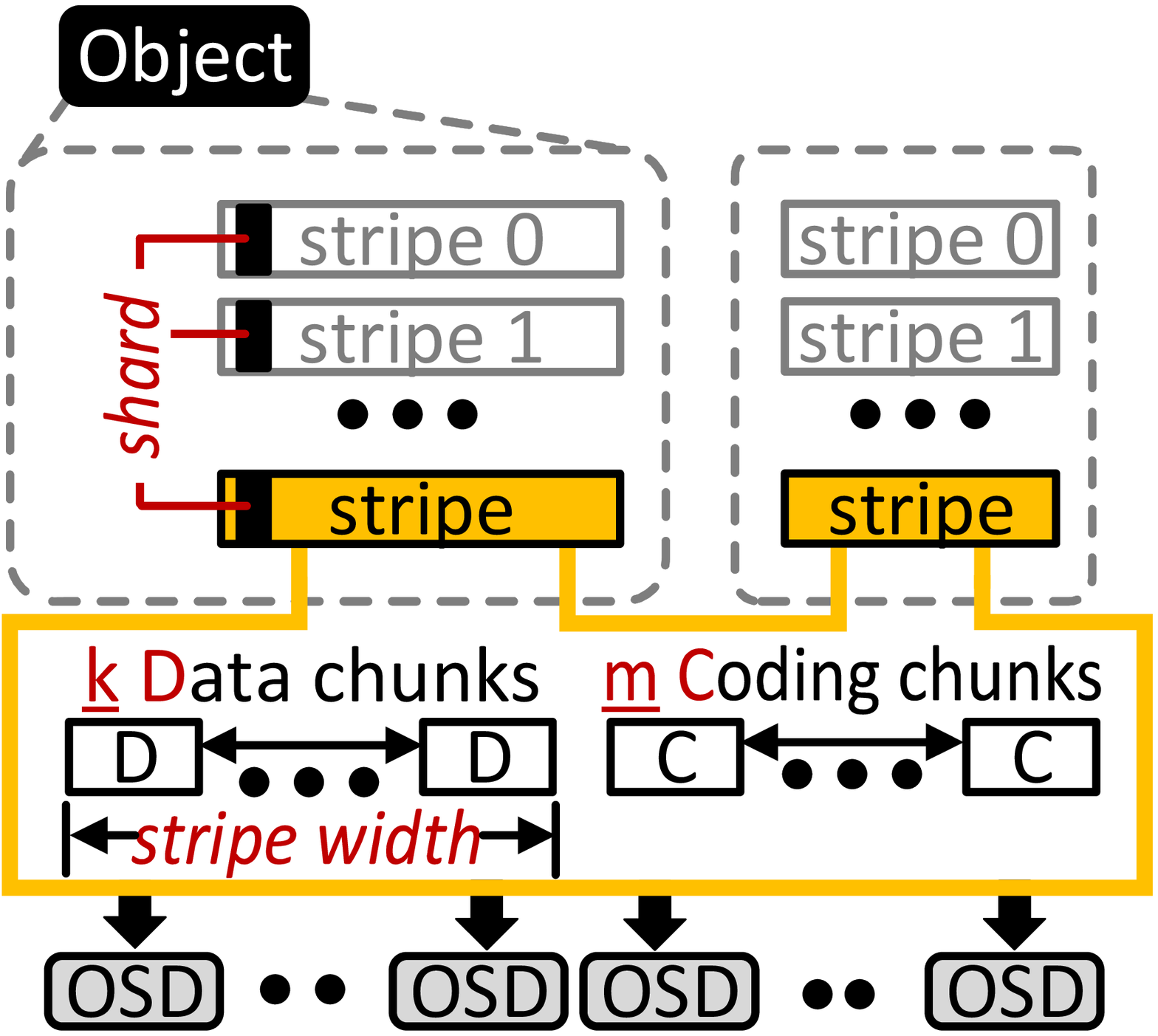}
\vspace{-10pt}
\caption{Terminology.}
\label{fig:term}
\end{subfigure}
~
\begin{subfigure}[b]{0.35\linewidth}
\includegraphics[width=1\linewidth]{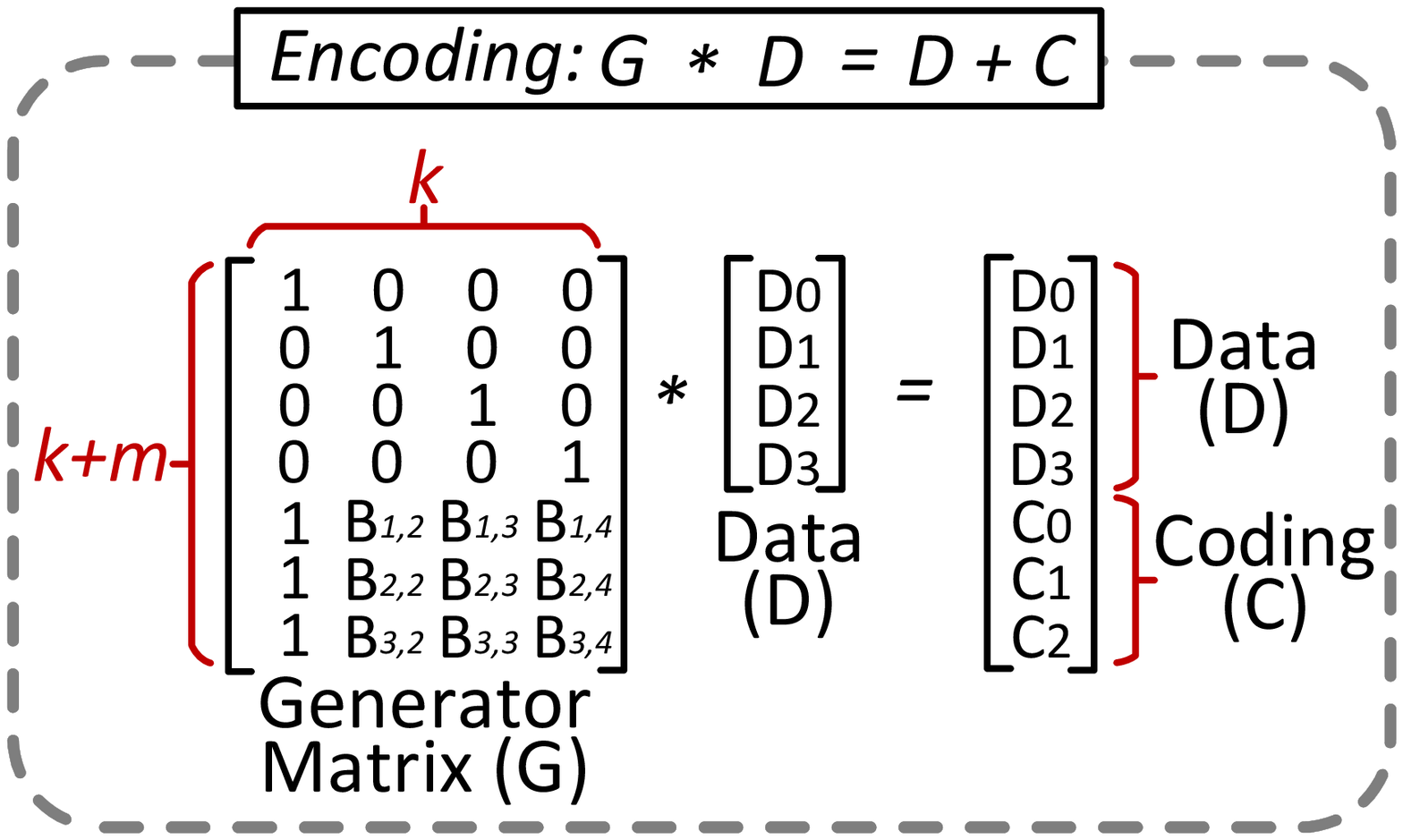}
\vspace{-10pt}
\caption{Encoding.}
\label{fig:encoding}
\end{subfigure}
~
\begin{subfigure}[b]{0.34\linewidth}
\includegraphics[width=1\linewidth]{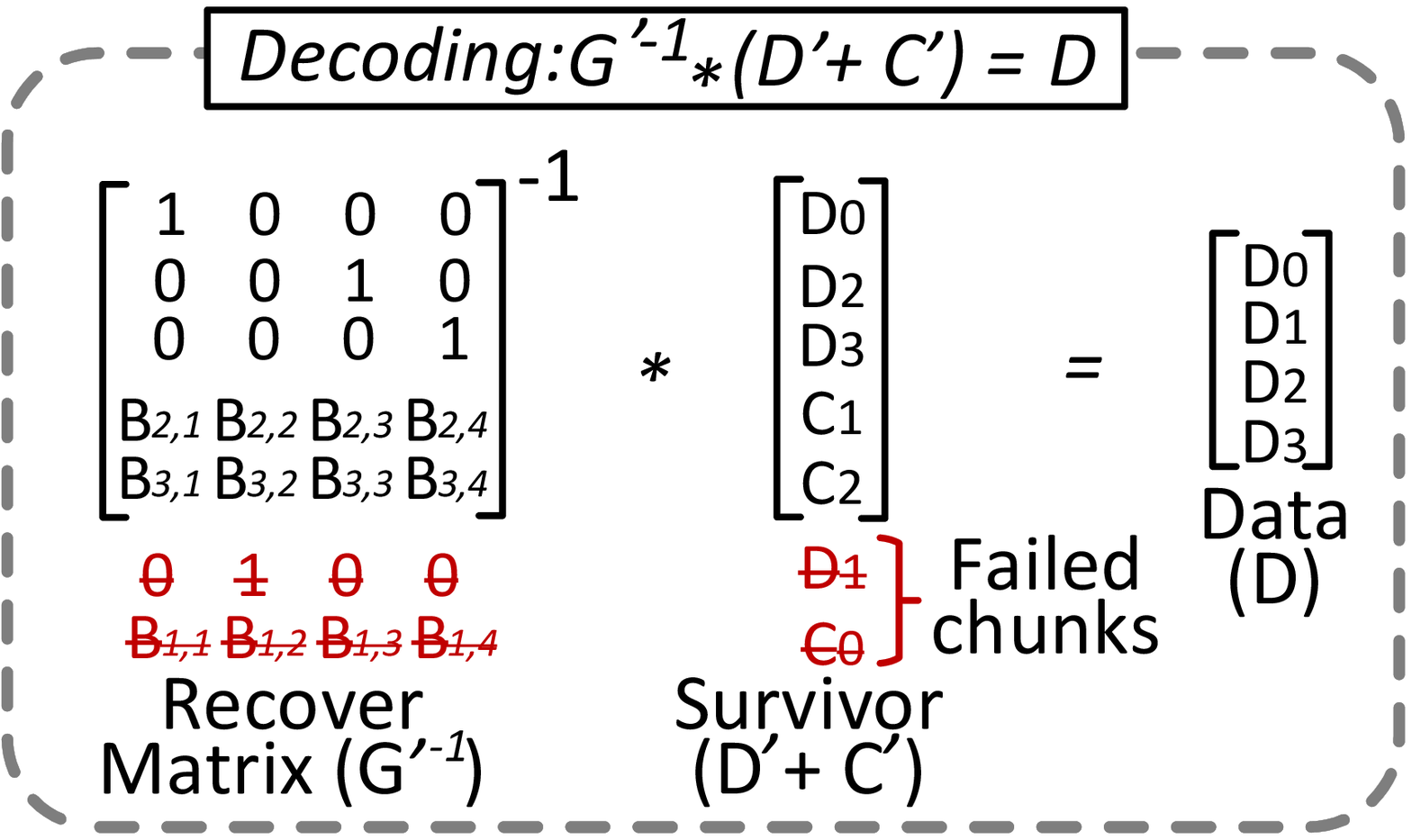}
\vspace{-10pt}
\caption{Decoding.}
\label{fig:decoding}
\end{subfigure}
\vspace{-9pt}
\caption{Erasure coding in Ceph parallel file system.\vspace{-17pt}}
\label{fig:overview}
\end{figure*}

\vspace{-5pt}
\subsection{Distributed SSD Array Systems and Ceph}
\vspace{-7pt}
\noindent\textbf{Overall architecture.} Figure \ref{fig:ceph} illustrates the architecture of a distributed SSD array system, which is managed by a Ceph parallel file system \cite{weil2006ceph}. At a high-level viewpoint, the client nodes and storage nodes are constructed by multiple server computers. Each client node can execute multiple host applications, and these applications can access the underlying storage cluster through a key library of Ceph, called \emph{reliable autonomic distributed object store} (RADOS). This RADOS library module (i.e., libRADOS) connects all underlying storage nodes as a cluster with client nodes. 
Since libRADOS manages the storage cluster via \emph{objects}, 
each including the corresponding data itself, a variable amount of metadata, and a globally unique identifier, RADOS block device (RBD) library (i.e., libRBD) is employed between libRADOS and applications that use a conventional block storage interface for their I/O services. On the other hand, the clients are connected to the storage nodes through a high speed network, referred to as \emph{public network}. Each storage node consists of several monitor daemons and one or more \emph{object storage device daemons} (OSDs). While an OSD handles read/write services from/to a storage device (i.e., an SSD in this work), a \emph{monitor} manages the layout of objects, the access permissions, and the status of multiple OSDs. The storage nodes are also connected together through a network, referred to as \emph{private network}, which is used for ensuring the reliability of underlying distributed storage systems. Through this network, multiple OSDs and monitors communicate with each other and check their daemons' health. For example, every six seconds, each OSD checks a heartbeat of other OSDs, thereby providing highly reliable and consistent backend data storage \cite{weil2007rados}.

\noindent \textbf{Data Path.} For an application running on a client, libRBD/libRADOS establishes a channel with the underlying storage cluster by retrieving a key and a cluster map from an underlying monitor, which includes the information about cluster topology related to mapping and authorization. Once the application sends an I/O request with the information of its target image and the corresponding address offset to libRBD, libRBD determines the object associated with the image and forward the request to libRADOS. This object is first mapped to one of the \emph{placement groups} (PGs) by determining an object ID with a simple hash function and adjusting the object ID aligned with the total number of PGs that the corresponding pool includes. After determining the corresponding PG ID, \emph{controlled replication under scalable hashing} (CRUSH) assigns an orderd list of OSDs to PG, based on the cluster topology and ruleset which represents the information of a given pool such as a type or the number of replicas \cite{weil2006crush}.
Lastly, libRADOS issues the I/O request to the primary OSD, which manages all the corresponding I/O services (usually related to resilience and workload balance) by referring all other OSDs listed in the PG map. This PG exists in the cluster map, which can be also retrieved by communicating with the monitor that resides at the primary OSD's node.

\noindent \textbf{Storage stack.} As shown in Figure \ref{fig:ceph}c, a client messenger is located at the top of the storage stack of each node, which handles all incoming I/O requests. Underneath the client messenger, a dispatcher fetches an I/O request and forwards it to the PG backend module. Since a storage node can fail at any given time with many unknown reasons, there is a log system that keeps track of all I/O requests fetched by the dispatcher, called PrimaryLogPG. The PG backend replicates data or encodes the data based on a given RS code. During this phase, it generates data copies (for replication) or data/coding chunks (for erasure coding), which are managed by a transaction module. This transaction module forwards the copies or chunks to another cluster messenger, which resides on the replica's or chunk's OSD. Note that, all these activities are performed by storage nodes without any intervention of client-side software. Especially, the data related to replication or erasure coding are transferred across multiple OSDs located in different storage nodes through the private network. Consequently, this data transfer is completely invisible to the client-side applications.

\vspace{-5pt}
\subsection{Fault Tolerance Mechanisms}
\vspace{-7pt}
When an application sends an I/O request to a specific target PG, it first acquires a lock, called \emph{PG lock}, to make sure that the data is successfully stored in the storage cluster. This lock can be released by a commit message that the corresponding primary OSD will return. During this phase, the primary OSD replicates data or performs erasure coding on the data to secure high resilience of the underlying storage cluster.

\vspace{-3pt}
\noindent \textbf{Replication.} Traditionally, a target system is made to be fault-tolerant by replicating data in parallel file systems \cite{weil2006ceph}. Ceph applies 3-replication as its default fault-tolerance mechanism \cite{cephpoolconfig}. Similar to a client that retrieves PG and the corresponding OSD through the cluster map (i.e., CRUSH and PG maps), any primary OSDs can examine the indices of OSD in PG through a copy of cluster map, which can be retrieved from the target node's monitor. The PG backend of the primary OSD writes replicated data (object) to the secondary OSD and the tertiary OSD through the transaction modules. If the target system adopts a stronger replication strategy, it can replicate the object to the appropriate PGs for the target as many OSDs as extra replicas. 
Once the process of replication is completed, the primary OSD informs the client that issued the I/O request of whether all the data are successfully stored by responding to a commit. Note that the amount of data that the primary OSD transmits through the private network can be at least 2$\times$ greater than that of data received from the public network.

\vspace{-7pt}
\noindent \textbf{Erasure coding.} Unfortunately, a replication method may not be straightforwardly applied to a storage cluster that employs only SSDs due to a high cost of SSDs per gigabyte. To address this, the distributed storage system community has paid attention to erasure coding since erasure coding introduces less storage overhead (less redundant information) for a given level of reliability that replication can offer. A very recent version of Ceph (version 11.2.0, released at 2017 January \cite{cephkrakenrelease}) has adopted
Read Solomon (RS) conding, which is one of the most popular and effective erasure coding techniques, and now is available for a block device interface (libRBD). RS coding is in practice classified as maximum distance separable (MDS) codes \cite{rosenthal1999maximum}, which is an optimal technique that secures the highest reliability within a given storage budget. As shown in Figure \ref{fig:term}, RS codes have two types of chunks to manage fault tolerance of $N$ size of target data: (1) \emph{data chunk} and (2) \emph{coding chunk} (also known as parity chunk). While $k$ data chunks are related to the original contents of target data, $m$ coding chunks maintain the parity data, which can be calculated from the $k$ data chunks. A \emph{stripe} is the unit that an RS code encode, and it consists of $k$ data chunks and $m$ coding chunks. The size of stripe, referred to as \emph{stripe width}, is $k\times n$ where $n$ is usually 4 KB in Ceph. 
Since failures can occur at both SSD-level and/or node-level, Ceph stripes the data chunks and coding chunks into $k+m$ different OSDs.
To make the binary data in an object reliable, Ceph manages data chunks and coding chunks of RS($k,m$) per object whose default size is 4 MB. Thus, for a given RS($k,m$) code, there are $N/(k*n)$ stripes, and all the chunks that have a same offset for the $N/(k*n)$ stripes are referred to as \emph{shard}, and all of them are managed by different OSDs.

\begin{figure}
\centering
\includegraphics[width=1\linewidth]{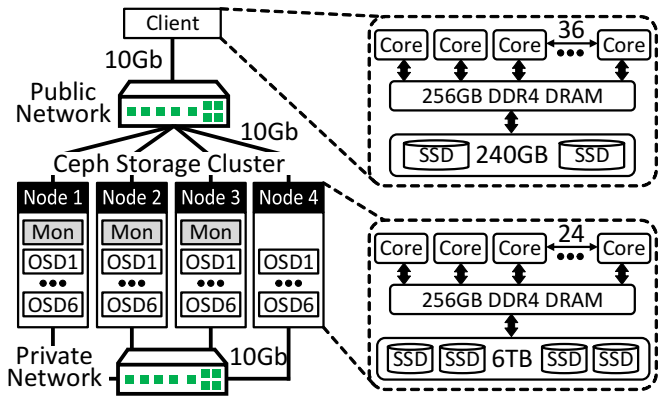}
\vspace{-20pt}
\caption{Configuration of our storage cluster.\vspace{-15pt}}
\label{fig:spec}
\end{figure}

\vspace{-9pt}
\subsection{Online Encoding and Repairs for a Failure}
\vspace{-9pt}
\noindent \textbf{Encoding.} Figure \ref{fig:encoding} shows a matrix-vector multiplication to generate $m$ coding chunks from $k$ data chunks. The $k$ words ($w$ bits of each data chunks) are considered as a vector, and $m$ coding chunks can be calculated by multiplying the vector with a matrix, referred to as \emph{generator matrix} (G). This G is $(k+m)*k$ geometric progression matrix, which is constructed by Vandermonde matrix \cite{lacan2004systematic}, and each element of the matrix is calculated by Galois Field arithmetic \cite{omura1986computational}. Each row of Vandermonde matrix has a form of a geometric sequence that begins with 1. $(k+m)*k$ \emph{extended Vandermonde matrix} is required in order to construct a generator matrix, whose first and last rows are same with that of $k*k$ \emph{identity matrix} respectively, while the rest of the matrix complies with the general form of the Vandermonde matrix. This extended matrix is converted to a generator matrix which has a same size with extended matrix, by multiplying a row/column by a constant and adding a row/column to another row/column. The first $k$ rows compose $k*k$ identity matrix and the following $m$ rows compose \emph{coding matrix} whose first row with all 1, as we can see in Figure \ref{fig:encoding}.

\noindent \textbf{Decoding.}
When the failure occurs \emph{recover matrix} is constructed, which is an inverse matrix of generator matrix whose rows that were calculated with failure chunks in encoding operation are removed. This inverse matrix is multiplied with remaining chunks, just the same way used at encoding operation. As a result of matrix-vector multiplication, we can get original chunks as we can see in Fiugre \ref{fig:decoding}. Due to the overheads greater than encoding of RS codes as we have to read remaining chunks and construct recover matrix, repairing bandwidth and performance degradation on decoding is a well-known problem \cite{huang2012erasure, sathiamoorthy2013xoring, esmaili2013core}. For example, Facebook uses RS(10,4) with 256MB chunk size, which generates 2GB data traffic to reconstruct data for a single data server \cite{sathiamoorthy2013xoring}. However, we observed that it is challenging to encode data online, which is not observed at prior studies. From the remaining sections, we will examine the performance impacts and overheads imposed by online encoding on an SSD array cluster.

%% file: evalsetup.tex
\vspace{-5pt}

In this section, we describe our SSD-based mini storage cluster and its system configuration.

\vspace{-5pt}

\noindent \textbf{Target node architecture.} Figure \ref{fig:spec} illustrates a real system that we built for the evaluation of performance impacts on erasure coding and the analysis of its system implications. The client employs 36 2.3 GHz cores (Intel Xeon E5-2669) and 256GB DDR4-2133 DRAM (16 channels, 2 ranks per channel). All operating system and executable images are booted from and store their local data to the client-side SSD. This client is connected to four different storage nodes of a Ceph storage cluster through a 10Gb public network. Each storage node also employs 24 2.6GHz cores (Intel Xeon E5-2690) and 256GB DDR4-2133 DRAM. For the storage, a high-performance SSD (600GB) is employed for the local OS image, while OSD daemons consist of 12 Intel SSD 730 (6TB), each grouping two Intel 730 through a hardware striping RAID controller \cite{ref-raid}. In total, there are 1TB DRAM, 96 cores, 52 SSDs (26TB) in the Ceph storage cluster that we built. The storage nodes are connected to another 10Gb (private) network, which is separated from the public network.  

\vspace{-5pt}

\noindent \textbf{Software and workload.}
For the parallel file system, we install Ceph 11.2.0 Kraken, which is the most recently released version (January 2017). Ceph Kraken we employed uses the Jerasure plugin module \cite{plank2014jerasure} and has Bluestore optimized for modern SSDs. To see the impact of the most common usage scenario of storage clusters, we use a flexible I/O tester, FIO \cite{axboe2015flexible}, on our mini cluster. Specifically, we create a data pool and a metadata pool separately, and measure overall system performance of the data pool by setting it up with 100 GB data per image. Each image contains 1024 PGs, and we collect the block traces for both data and metadata pools using blktrace \cite{brunelle2006block}. We performed pre-evaluation to see the number of queue entries that exhibit the best performance of the underlying storage cluster. We observed that the queue depth of 256 is the best for diverse workloads, and therefore applied the queue depth to all evaluation scenarios performed by this work. When there is no data written upon the target storage, the read performance can be far from the actual storage performance since the underlying SSD in practice just serves it with garbage values from DRAM if there is no target data written (rather than accessing actual Flash media). Thus, we sequentially write 4MB data to the target image before performing each set of evaluations. These pre-write operations also are applied to overwrite tests in Section \ref{sec:distribution}. Lastly, it is possible to have a performance impact due to physical data layout or fragment of the written data, we deleted pool and image and created them again before the test begins for each evaluation.
.

%% file: overall.tex
\vspace{-5pt}

In this section, we compare the performance of two major erasure coding configurations, RS(6,3) and RS(10,4), with that of 3-replication at a high level.

\begin{figure}
	\centering
	\begin{subfigure}{0.48\columnwidth}
		\includegraphics[width=\columnwidth]{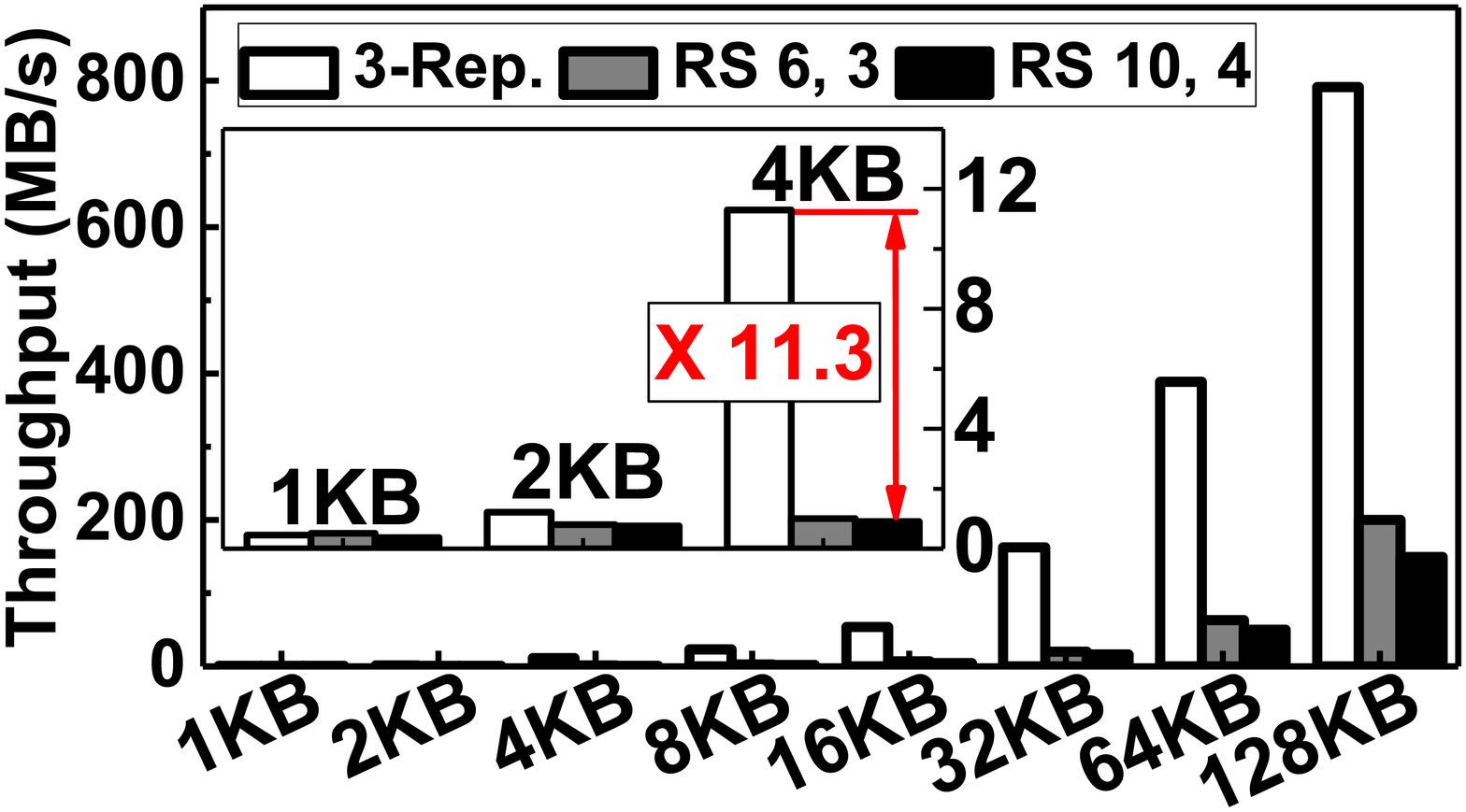}
		\caption{Throughput.}
		\label{fig:seq_write_throughput}
	\end{subfigure}
	~
	\begin{subfigure}{0.48\columnwidth}
		\includegraphics[width=\columnwidth]{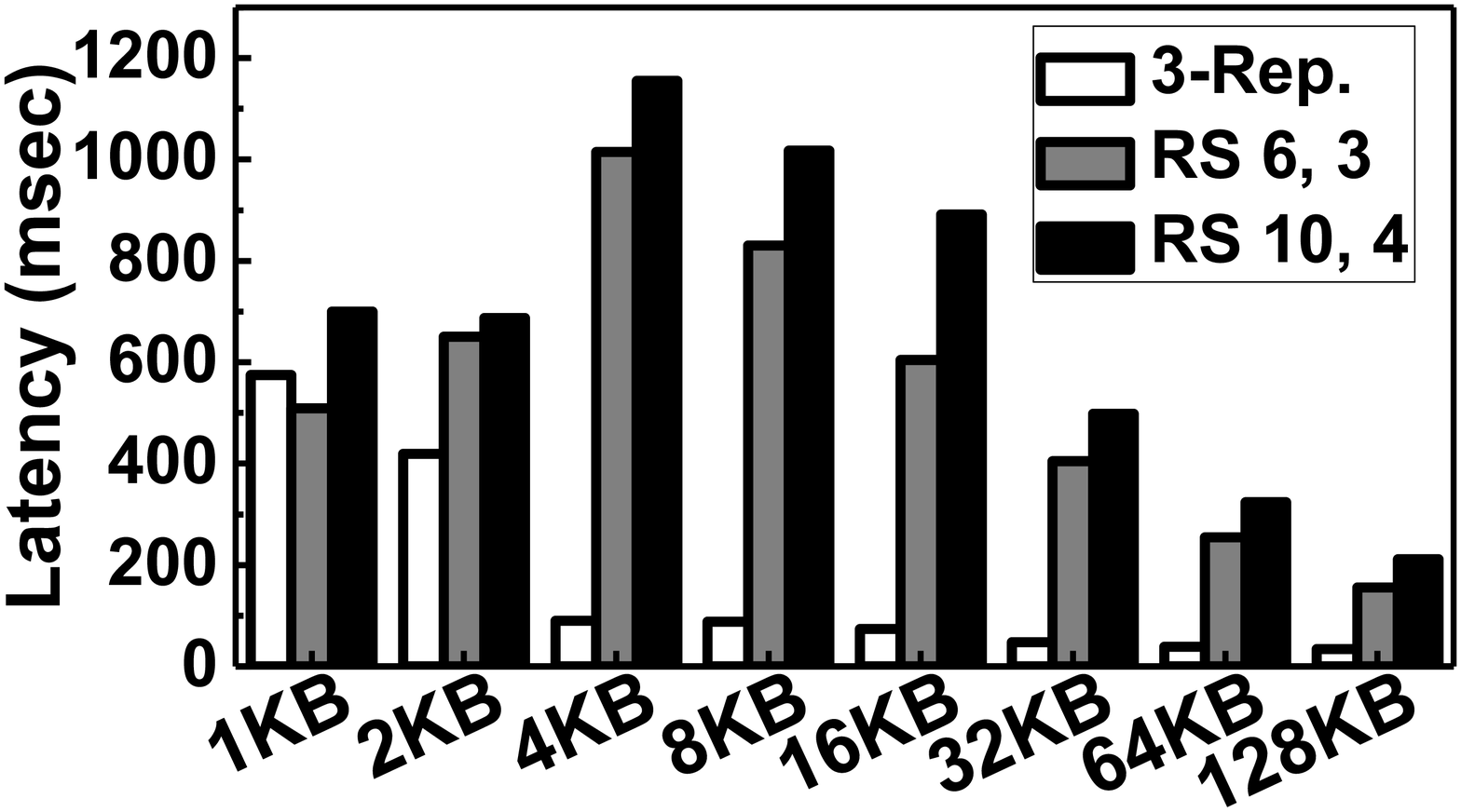}
		\caption{Latency.}
		\label{fig:seq_write_latency}
	\end{subfigure}
	\vspace{-6pt}
	\caption{Comparison of sequential write performance. \vspace{-11pt}}
	\label{fig:seq_write_perf}
\end{figure}

\begin{figure}
	\centering
	\begin{subfigure}{0.48\columnwidth}
		\includegraphics[width=\columnwidth]{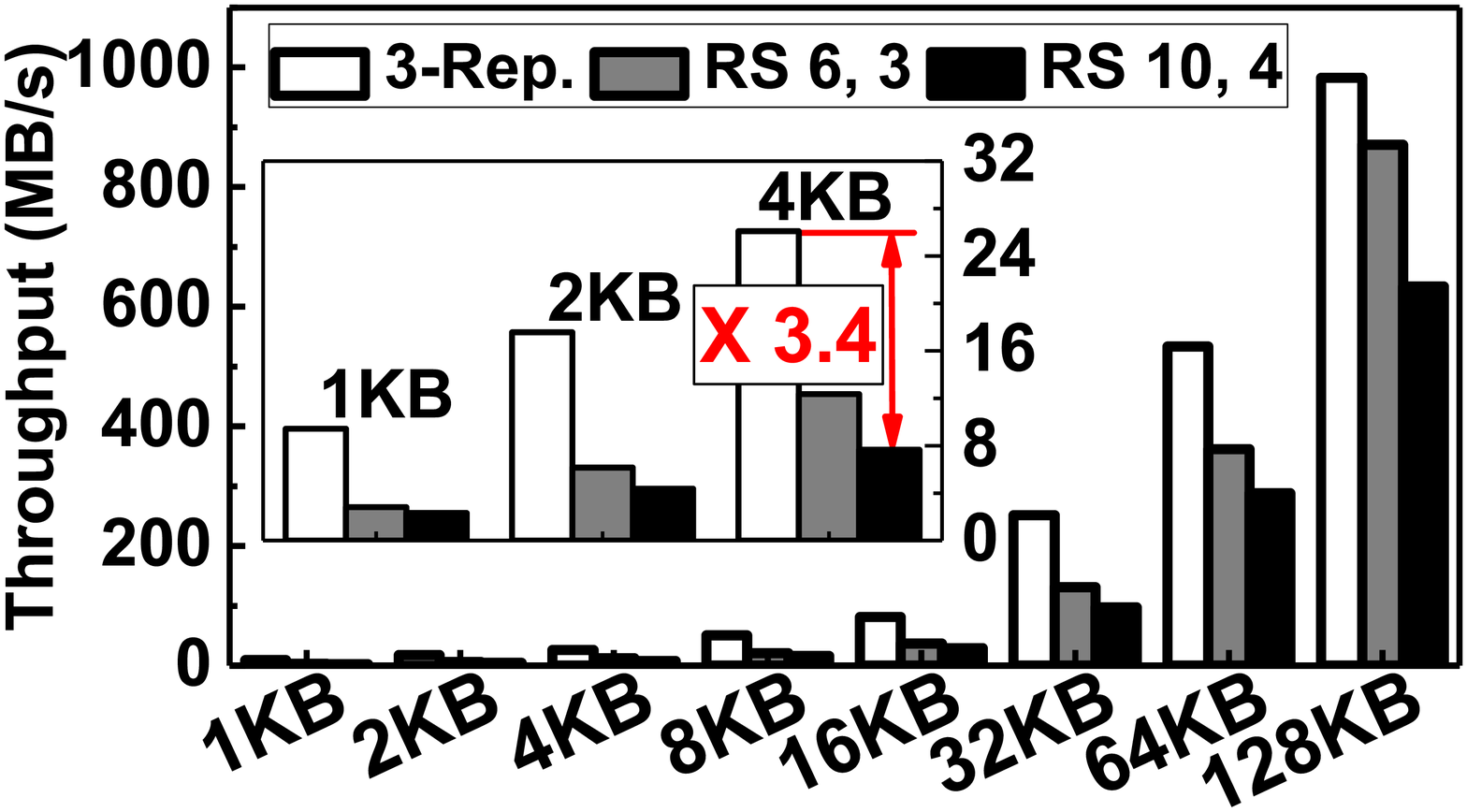}
		\caption{Throughput.}
		\label{fig:seq_read_throughput}
	\end{subfigure}
	~
	\begin{subfigure}{0.48\columnwidth}
		\includegraphics[width=\columnwidth]{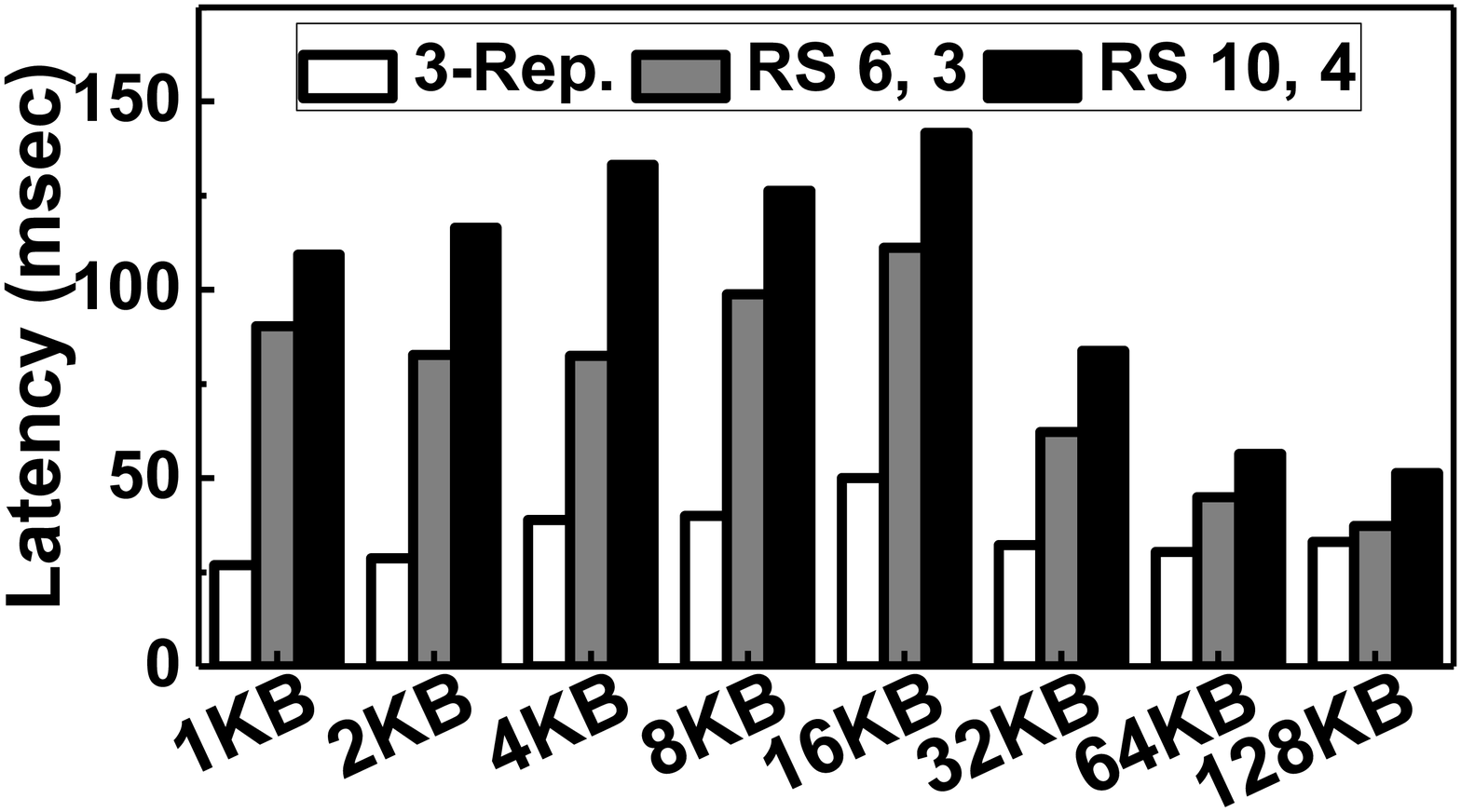}
		\caption{Latency.}
		\label{fig:seq_read_latency}
	\end{subfigure}
	\vspace{-9pt}
	\caption{Comparison of sequential read performance. \vspace{-18pt}}
	\label{fig:seq_read_perf}
\end{figure}

\vspace{-5pt}

\subsection{Sequential Performance}
\vspace{-5pt}
\noindent \textbf{Writes.} Figures \ref{fig:seq_write_throughput} and \ref{fig:seq_write_latency} show the throughput and latency of sequential writes with various block sizes ranging from 1KB to 128KB, respectively. As shown in Figure \ref{fig:seq_write_throughput}, 3-replication offers around 179 MB/s on average, whereas RS(6,3) and RS(10,4) give 36.8 MB/s and 28.0 MB/s for the sequential writes, respectively. Overall, the throughput of RS(6,3) is worse than 3-replication by 8.6$\times$ for 4KB$\sim$16KB request sizes. Considering that the most popular request size of diverse applications and native file systems is 4KB$\sim$16KB, the performance degradation of online RS coding may not be acceptable in many computing domains. As shown in Figure \ref{fig:seq_write_latency}, the latency of RS(6,3) is 3.2$\times$ longer than that of 3-replication on average. Note that the latency of a conventional Ceph configuration (with 3-replication) is less than 90 ms for most block sizes of I/O accesses. We believe that the latency of 3-replication is in a reasonable range. However, RS(6,3) and RS(10,4) require 544 ms and 683 ms, respectively, on average, 
Such long latencies can be a serious issue for many latency-sensitive applications. The reason behind the long latency and low bandwidth of RS(6,3) and RS(10,4) is that online erasure coding requires computation for encoding, data management, and additional network traffic. We will analyze each contribution of performance degradation in more detail in Section \ref{sec:system} and \ref{sec:dataoverhead}. Note that the latency on sequential writes becomes shorter as the block size increases. This is because the consecutive requests can be delayed by PG backend until the service of previous request is completed while a small size of request reads actually out the data as stripe width. 
We will examine this phenomenon in details in Section \ref{sec:distribution}. 

\noindent \textbf{Reads.} Figure \ref{fig:seq_read_perf} compares the throughput of RS(6,3) and RS(10,4) with that of 3-replication by performing sequential read operations. In contrast to the writes (cf. Figure 5), RS(6,3) and RS(10,4) degrade the performance by only 26\% and 45\% on average, compared with 3-replication. This is because decoding process (i.e., repair bandwidth) is only involved when there is a device or node failure. Nevertheless, RS(6,3) and RS(10,4) yet give 2.2$\times$ and 2.9$\times$ longer latency than that of 3-replication, respectively. Even though the decoding and repair overheads occur only when there is a failure, the reads of online erasure coding requires composing the data chunks into a stripe, which in turn introduces the overheads not observed by any replication method. We refer this process associated with reads to as \emph{RS-concatenation}, which also generates extra data transfers over the private network, thereby increasing the latency of RS(6,3) and RS(10,4) further. As shown in Figure \ref{fig:seq_read_latency}, the latency of traditional configuration (with 3-replication) is less than 50 ms, whereas RS(6,3) and RS(10,4) require as high as 111 ms and 142 ms, respectively. We will examine the performance and system impacts of RS-concatenation in Section \ref{sec:system}.

\begin{figure}
	\centering
	\label{fig:seq_write_perf}
	\begin{subfigure}{0.48\columnwidth}
		\includegraphics[width=\columnwidth]{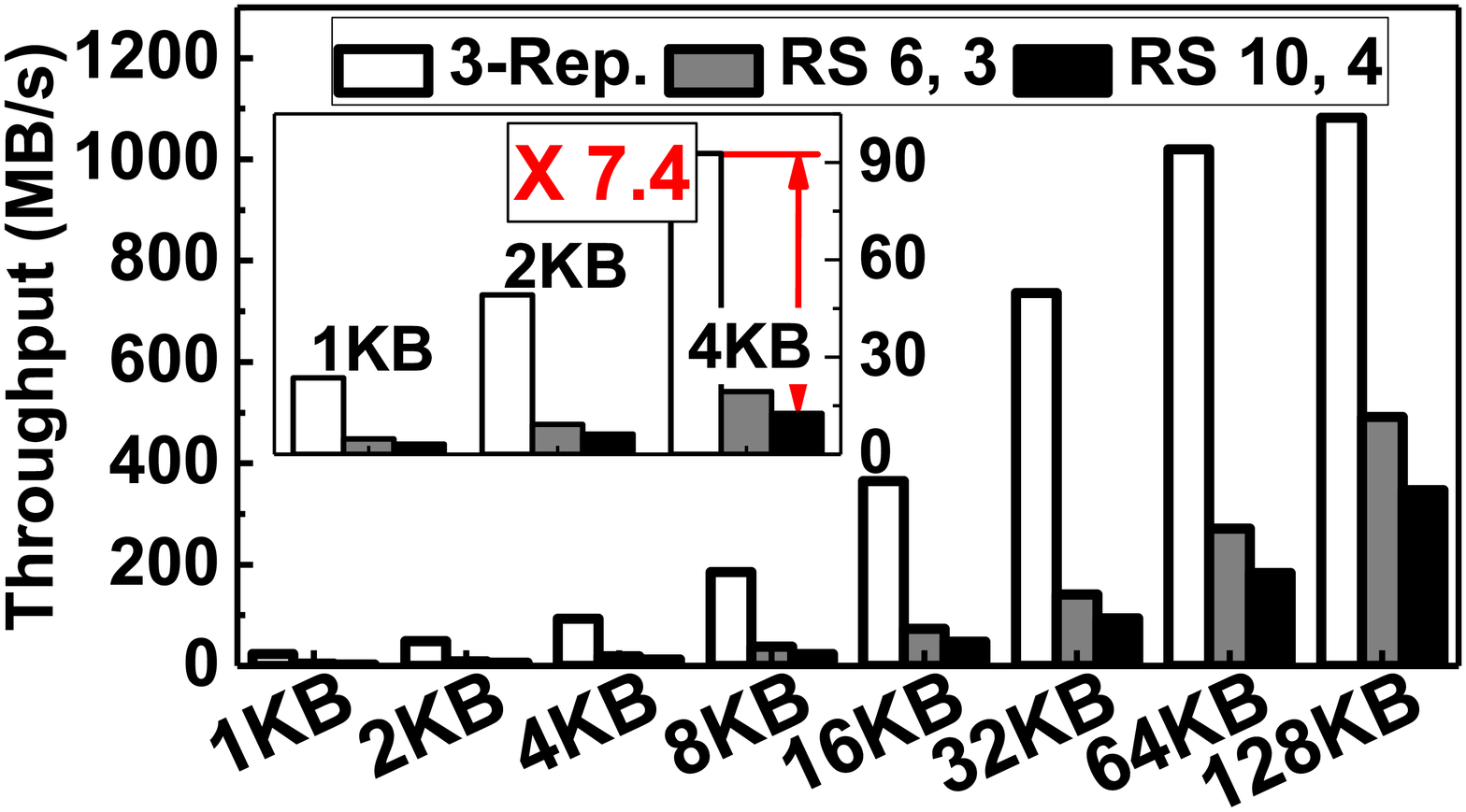}	
		\caption{Throughput.}
		\label{fig:rand_write_throughput}
	\end{subfigure}
	~
	\begin{subfigure}{0.48\columnwidth}
		\includegraphics[width=\columnwidth]{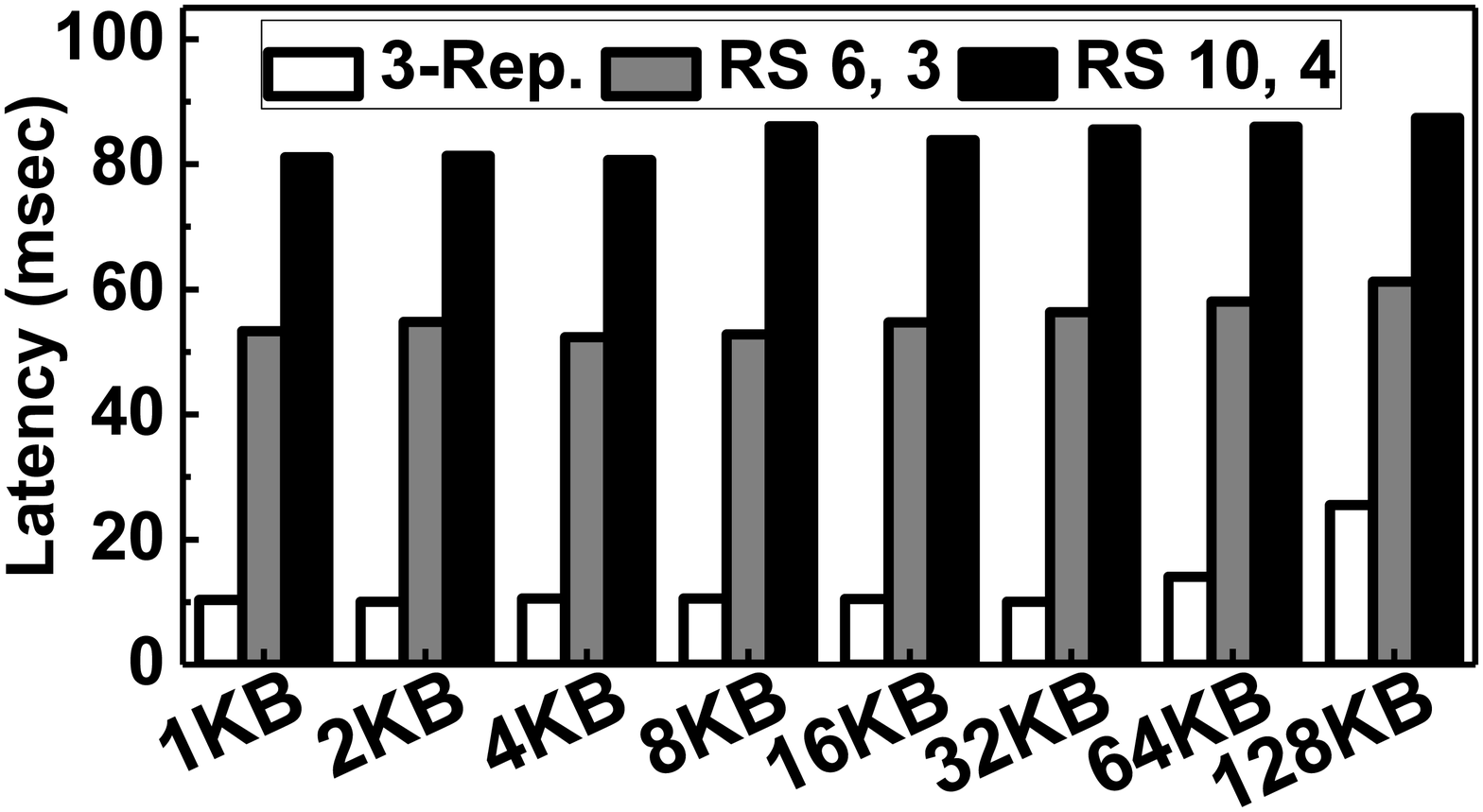}	
		\caption{Latency.}
		\label{fig:rand_write_latency}
	\end{subfigure}
	\vspace{-6pt}
	\caption{Comparison of random write performance. \vspace{-11pt}}
	\label{fig:rand_write_perf}
\end{figure}

\begin{figure}
	\centering
	\begin{subfigure}{0.48\columnwidth}
		\includegraphics[width=\columnwidth]{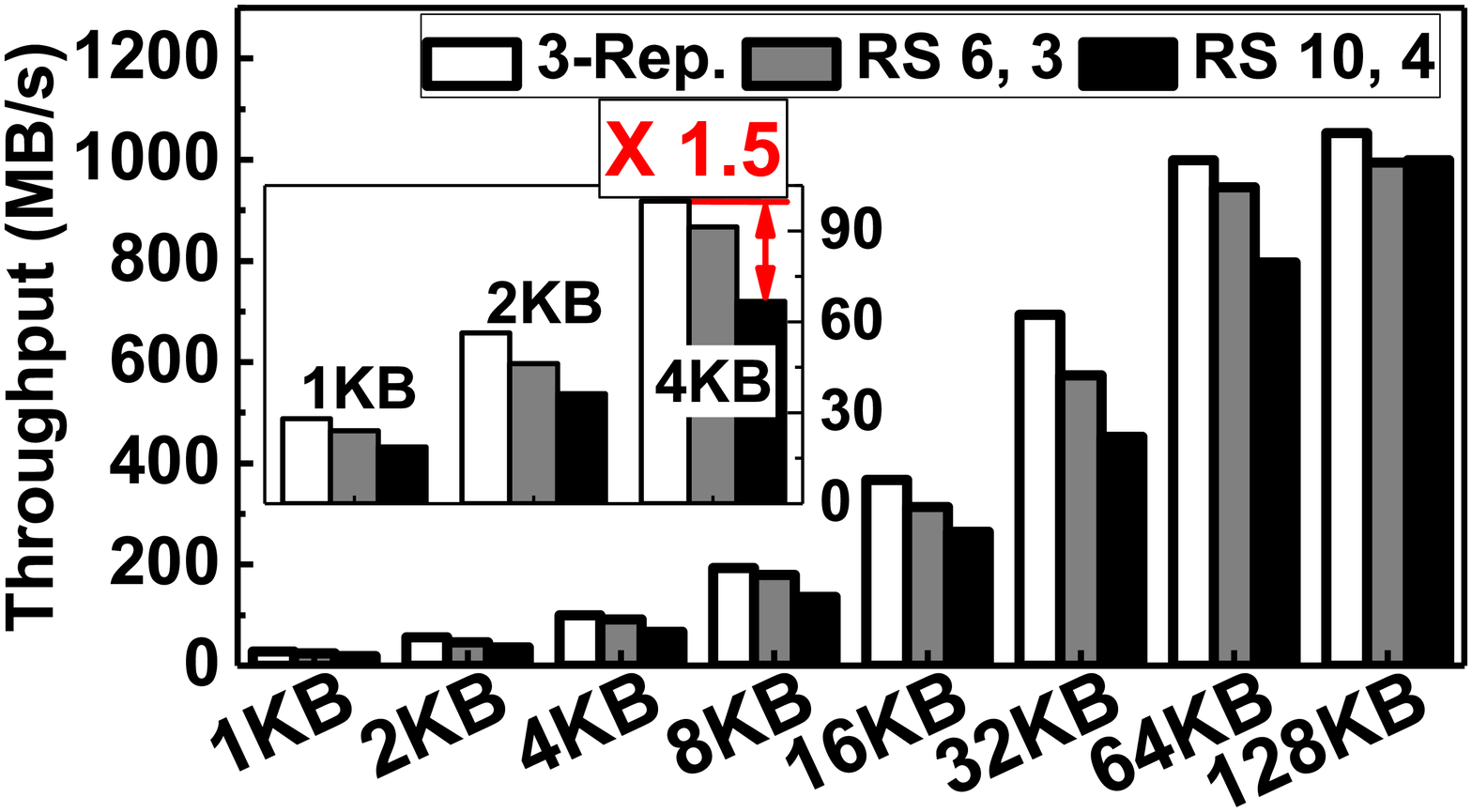}
		\caption{Throughput}
		\label{fig:rand_read_throughput}
	\end{subfigure}
	~
	\begin{subfigure}{0.48\columnwidth}
		\includegraphics[width=\columnwidth]{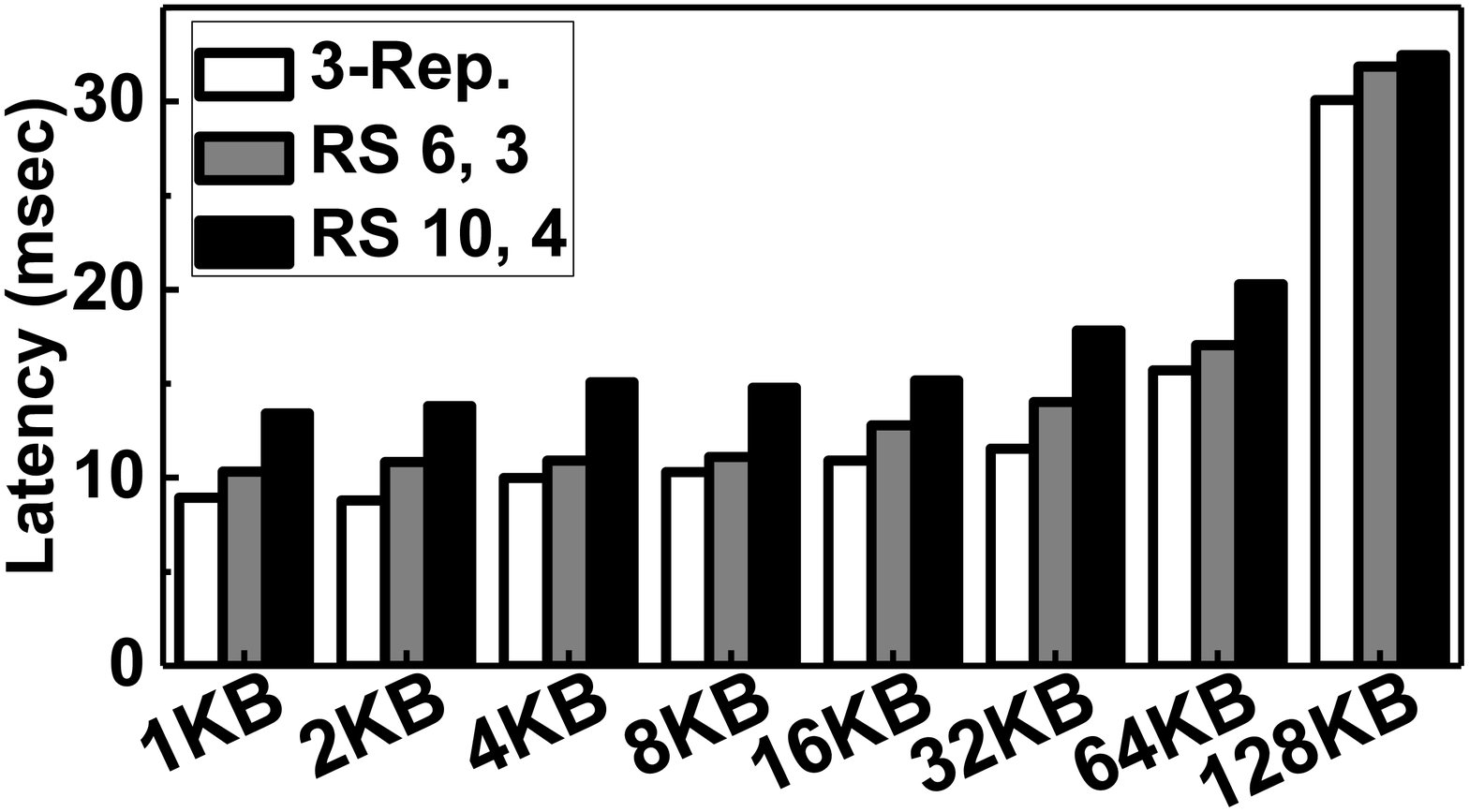}
		\caption{Latency}
		\label{fig:rand_read_latency}
	\end{subfigure}
	\vspace{-9pt}
	\caption{Comparison of random read performance. \vspace{-18pt}}
	\label{fig:rand_read_perf}
\end{figure}

\vspace{-10pt}
\subsection{Random Performance}
\vspace{-8pt}
\noindent \textbf{Writes.} As shown in Figure \ref{fig:rand_write_perf}, RS(6,3) and RS(10,4) give 3.4$\times$ and 4.9$\times$ worse write performance than 3-replication for random I/O accesses, respectively. The trend of performance difference is similar to that of sequential writes, but the random write bandwidth of RS(6,3) and RS(10,4) offer 3.6$\times$ and 3.2$\times$ higher write bandwidth for random writes than for sequential writes, respectively. This is because the sequential accesses with the block size smaller than an object mostly target to the same PG at its primary OSD, which makes many resource conflicts at the PG level. As explained in Section \ref{sec:background}, since the target's dispatcher (underneath its client messenger) locks the target PG to offer strong consistency of a storage cluster, the underlying PG backend is not available, which renders lower throughput and exhibits longer latency. In contrast, since I/O requests under random accesses can be distributed across different OSDs (in our case 24 OSDs), such kind of lock contentions can be addressed at some extent, which in turn increases performance. Note that this phenomenon is also observed in 3-replication, but the number of OSDs (and nodes) that the PG backend needs to handle through its cluster messenger are less than that of online erasure coding. 
This is also reason why the latency trend on random writes is completely different from that on sequential writes. While the latency of RS(6,3) and RS(10,4) becomes shorter as the block size increases, the latency of RS coding on random writes is sustainable, irrespective of the block size. This is because erasure coding of the PG backend performs I/O services based on stripe granularity (and the data chunks are distributed across different nodes), while 3-replication issues the request as much as it needs since all data can be served by a single OSD. 
More importantly, RS(6,3) and RS(10,4) provide 90\% and 88\% shorter latency for random writes than for sequential writes, respectively, on average.

\begin{figure}
	\centering
	\begin{subfigure}{0.48\columnwidth}
		\includegraphics[width=\columnwidth]{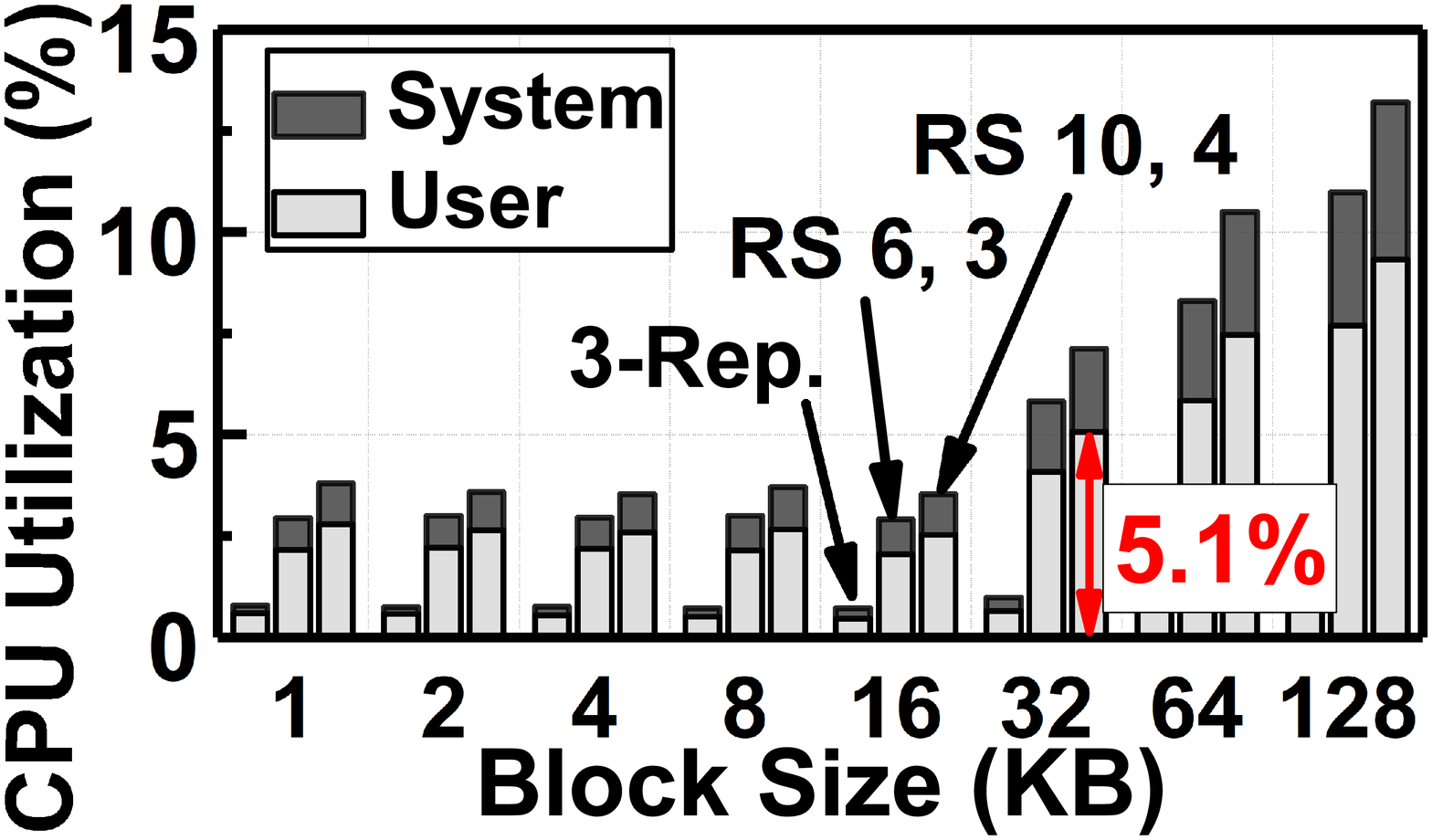}
		\caption{Sequential write.}
		\label{fig:seq_write_cpu}
	\end{subfigure}
	~
	\begin{subfigure}{0.48\columnwidth}
		\includegraphics[width=\columnwidth]{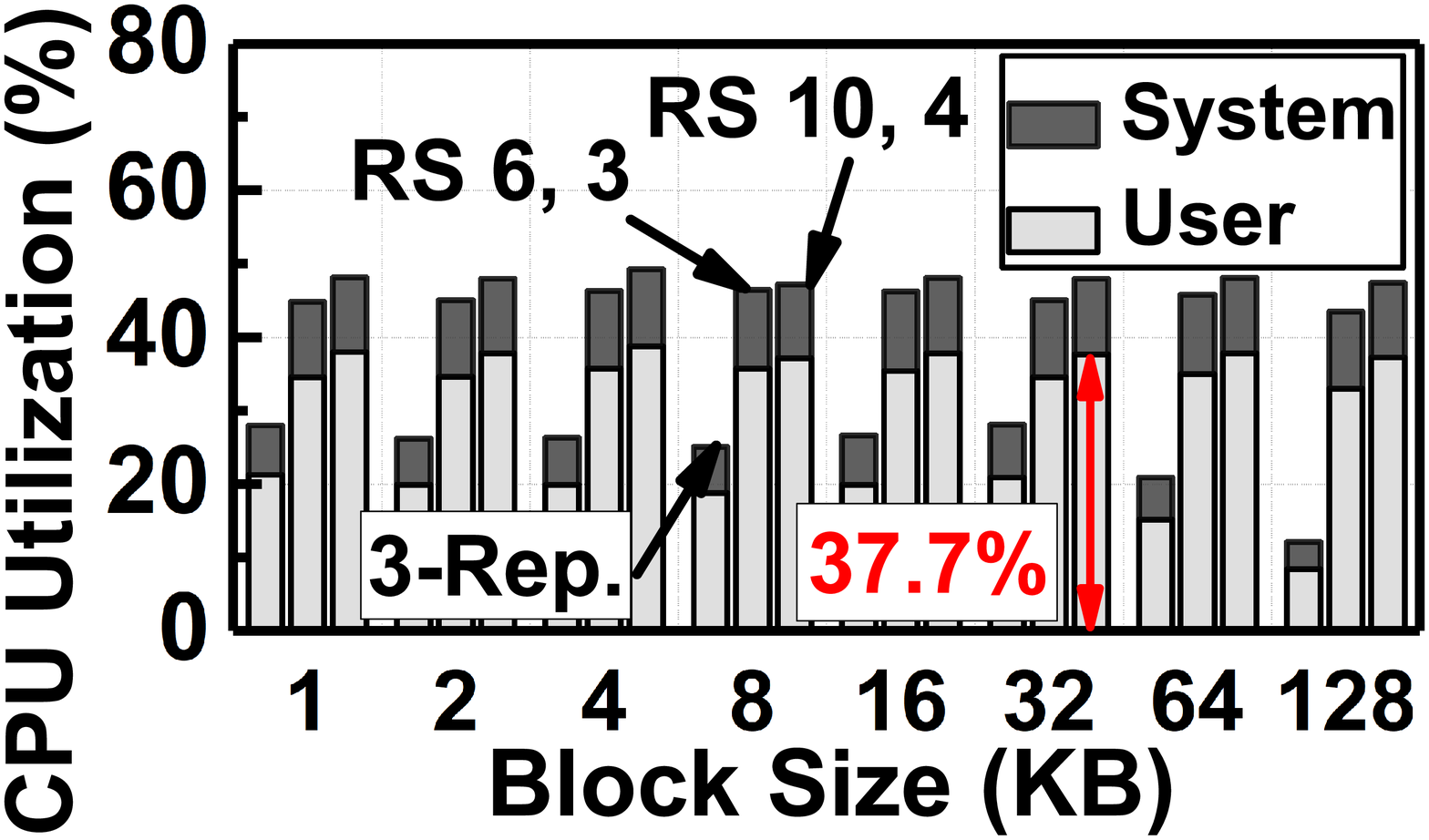}
		\caption{Random write.}
		\label{fig:rand_write_cpu}
	\end{subfigure}
	\vspace{-5pt}
	\caption{Average CPU utilization by writes. \vspace{-18pt}}
	\label{fig:avg_cpu_write}	
\end{figure}

\vspace{-5pt}
\noindent \textbf{Reads.} Figure \ref{fig:rand_read_perf} shows the read performance of 3-replication, RS(6,3) and RS(10,4) with random I/O accesses. In contrast to sequential reads, the difference between 3-replication and RS(6,3) in terms of bandwidth and latency is less than 10\% on average. RS-concatenation, which is a process to compose data chunks into a stripe, not only wastes computation but also requires pulling the data chunks from the underlying other OSDs. Since the primary OSD does not send the acknowledge signal to the client until all data chunks are arrived from the other RS-associated OSDs, the target read request can be delayed as long each request as delayed. While the impact of this overhead on performance is in a reasonable range, 
due to the similar reason to sequential writes, the sequential reads are more delayed because of the PG lock contention, which is not observed in random I/O accesses. 
Because of this higher degree of PG-level parallelism, the performance of random reads is also better than that of sequential reads. We will closely examine these performance issues on random accesses further in Section \ref{sec:distribution}.

%% file: system.tex
In this section, we will analyze the computing overheads (in terms of CPU utilizations and context switches) imposed by 3-replication, RS(6,3) and RS(10,4). To be precise, we performed pre-evaluation to measure the computing overheads involved with serving I/O subsystems in even when the system is idle and exclude such overheads from our results. The computing overheads we will report in this section are average values of 96 cores that our mini storage cluster employs. Note that the stripe width of RS(6,3) and RS(10,4) is 24KB and 40KB, respectively.

\begin{figure}
	\centering
	\begin{subfigure}{0.48\columnwidth}
		\includegraphics[width=\columnwidth]{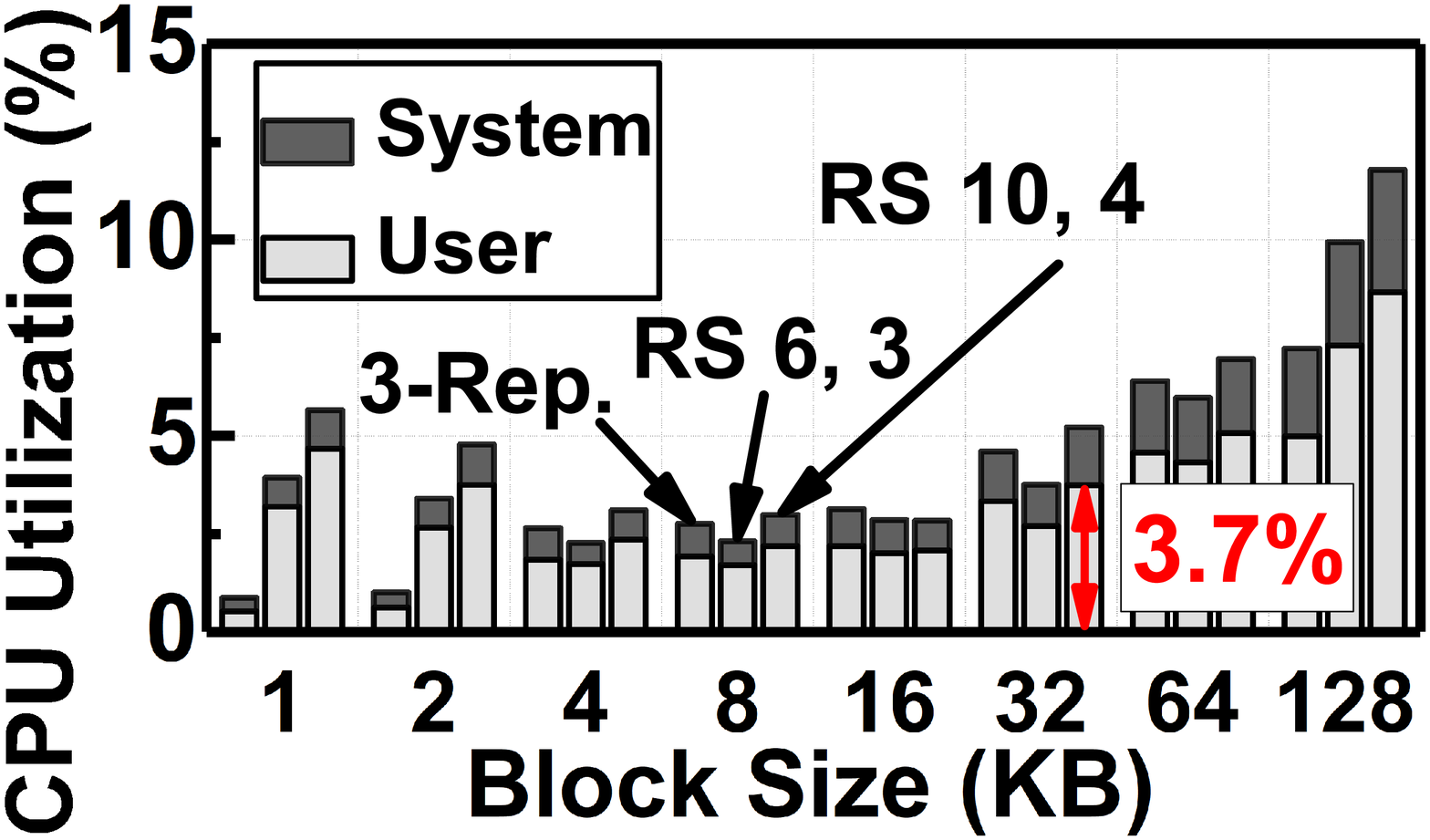}
		\caption{Sequential read.}
		\label{fig:seq_read_cpu}
	\end{subfigure}
	~
	\begin{subfigure}{0.48\columnwidth}
		\includegraphics[width=\columnwidth]{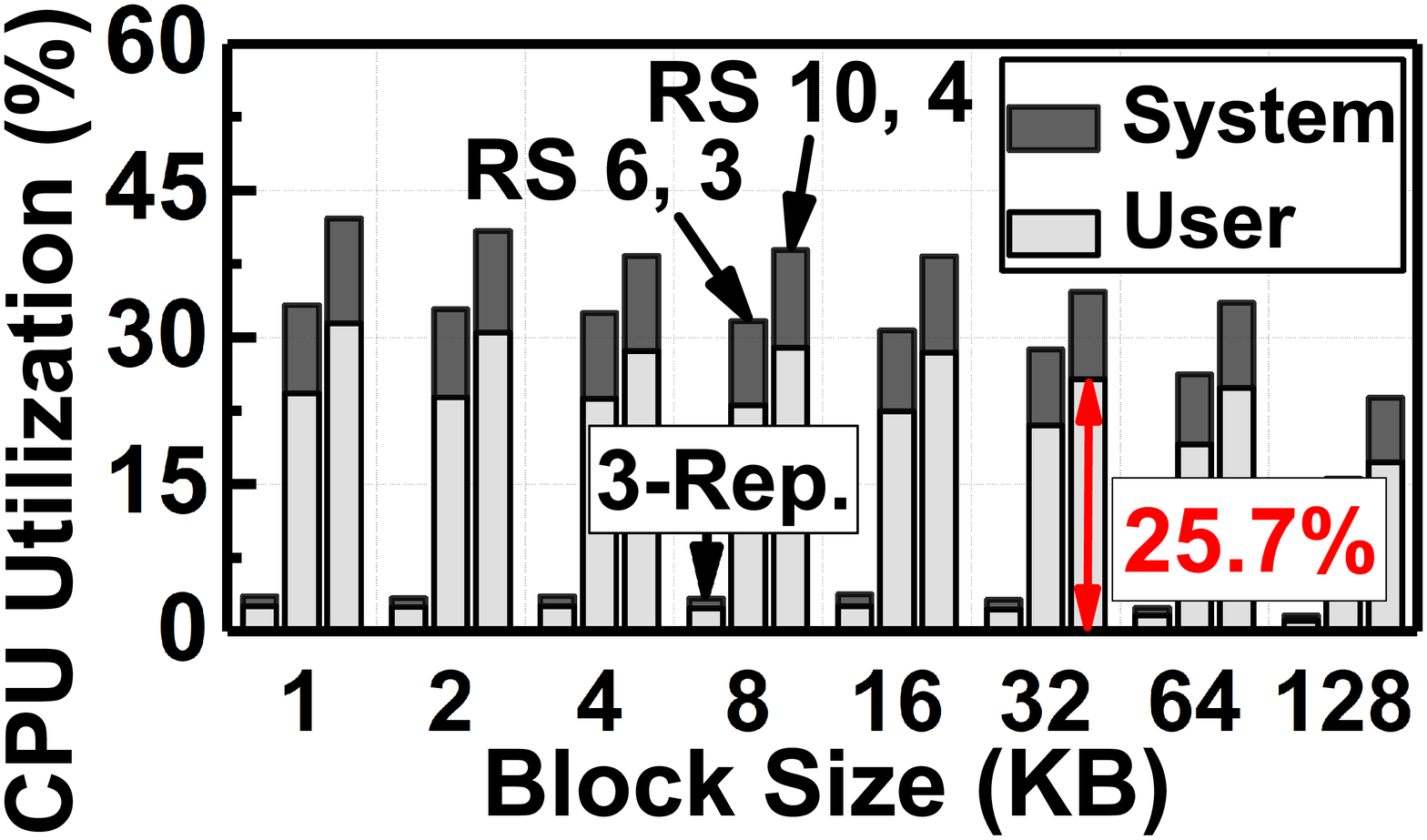}
		\caption{Random read.}
		\label{fig:rand_read_cpu}
	\end{subfigure}
	\vspace{-5pt}
	\caption{Average CPU utilization by reads. \vspace{-18pt}}
	\label{fig:avg_cpu_read}
\end{figure}

\vspace{-8pt}
\subsection{CPU Utilization}
\vspace{-8pt}
\noindent \textbf{Writes.} Figure \ref{fig:avg_cpu_write} shows the CPU utilization of 3-replication, RS(6,3) and RS(10,4) by writes. In these CPU utilization tests, we measure the utilization of kernel side (cf. system) and user side (cf. user) separately. For sequential writes, both replication and erasure coding consume around 4.4\% of total CPU execution cycles on average. 
A notable part of this measurement is that the user-mode CPU utilizations account for 70\%$\sim$75\% of total CPU cycles (storage cluster's 96 cores), which is not observed in conventional multiple driver layers in Linux. Since all OSD daemons including the PG backend and fault tolerance modules (including erasure coding and 3-replication) are implemented at the user level, user-mode operations take more CPU cycles than kernel-mode operations to secure storage resilience. On the other hand, as we can see from Figure \ref{fig:rand_write_cpu}, RS(6,3) and RS(10,4) requires 45\% and 48\% of total CPU cycles, which are about 2$\times$ greater than the CPU usages that 3-replication needs (24\% of total CPU cycles). 
The reason behind higher consumptions of CPU cycles on random accesses is that the amount of I/O requests arrived at PG backend is insufficient for processing both replication and erasure coding due to the PG locks which causes conflicts on the primary OSDs that we described in Section \ref{sec:background}.

\vspace{-5pt}
\noindent \textbf{Reads.} In contrast to writes, reads do not require encoding and/or decoding processes as there is no failure in our evaluation. However, we observe that CPU cycles are wasted by the fault tolerance modules of the PG backend in storage nodes. Figure \ref{fig:seq_read_cpu} shows the CPU cycles consumed by 3-replication, RS(6,3) and RS(10,4) under sequential reads. As we expected, 3-replication uses only 0.9\% of total CPU cycles, but RS(6,3) and RS(10,4) consume up to 5.0\% and 6.1\% CPU cycles, respectively. This is because, while 3-replication does not need to manipulate replicas to serve block I/O requests, erasure coding should concatenate data chunks distributed across multiple OSDs and compose them into a stripe. This RS-concatenation consumes CPU cycles not only for the stripe composition but also for transaction module to handle data transfers. This phenomenon is more prominent when serving random I/O requests. As shown in Figure \ref{fig:rand_read_cpu}, 3-replication only takes 3.1\% of CPU cycles, whereas RS(6,3) and RS(10,4) consumes 29.0\% and 36.3\% of total CPU cycles, respectively, on average. The reason why erasure coding requires more CPU cycles on random accesses is the same as what we described for random writes (i.e., PG locks and resource conflicts on primary OSDs). Note that, without even a failure or a decoding operation, erasure coding consumes 32.6\% of total CPU cycles, and user-mode operations (73.6\% of total CPU cycles) are always involved with it during just serving read requests, which are not acceptable in terms of power, efficiency and scalability in many computing domains.

\begin{figure}
	\centering
	\begin{subfigure}{0.48\columnwidth}
		\includegraphics[width=\columnwidth]{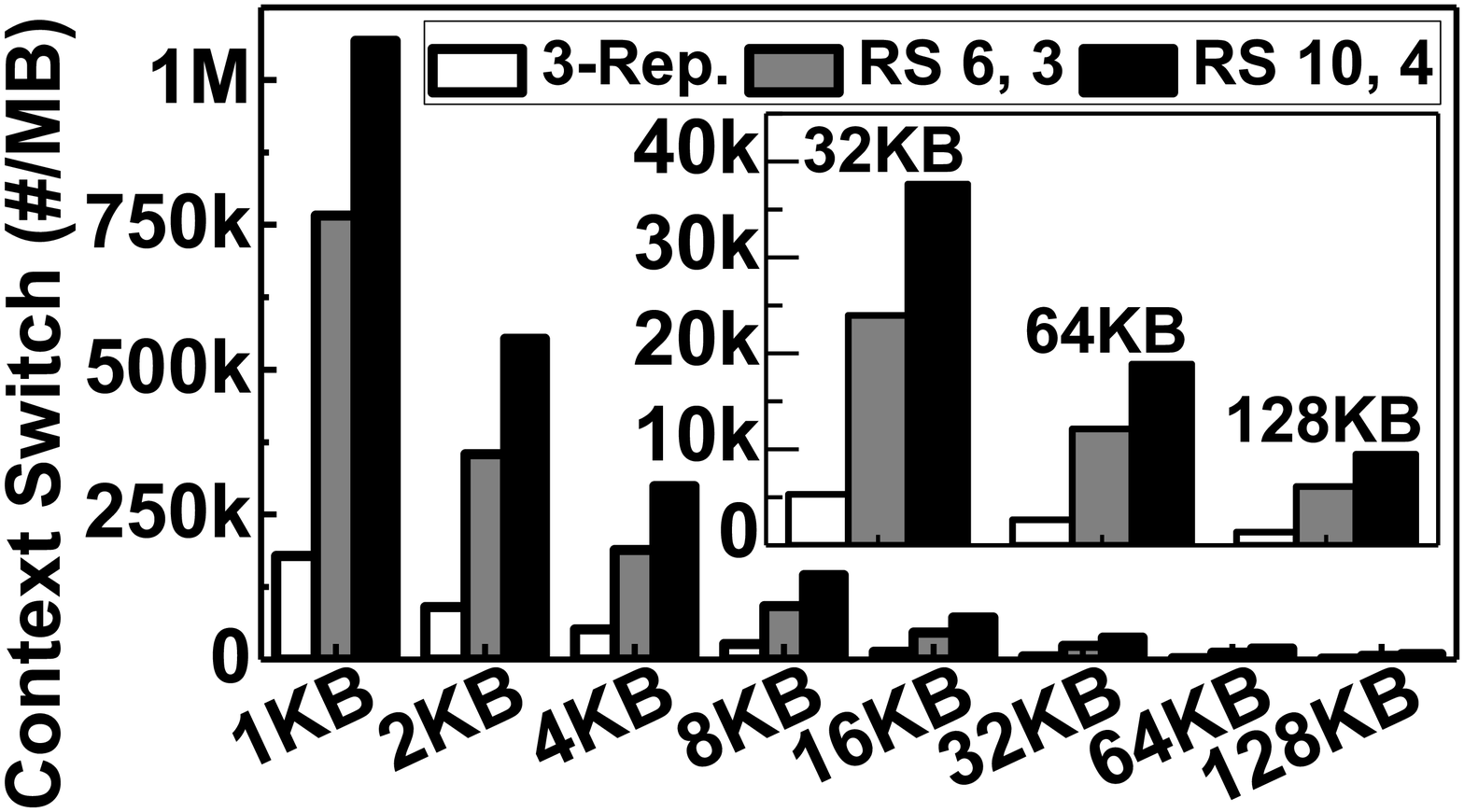}
		\caption{Sequential write.}
		\label{fig:rel_ctx_seq_write}
	\end{subfigure}
	~
	\begin{subfigure}{0.48\columnwidth}
		\includegraphics[width=\columnwidth]{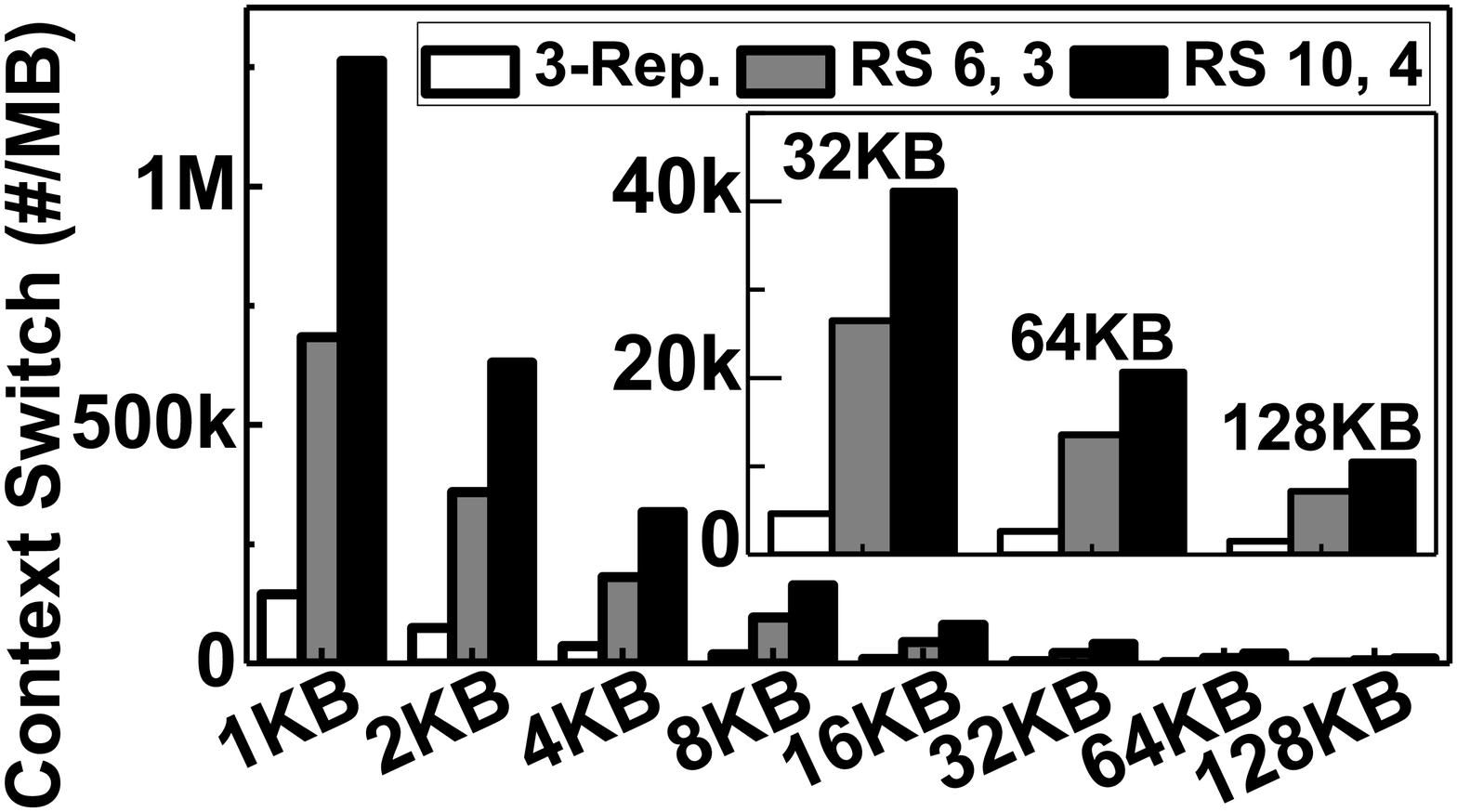}
		\caption{Random write.}
		\label{fig:rel_ctx_rand_write}
	\end{subfigure}
	\vspace{-7pt}
	\caption{relative number of context switches (writes). \vspace{-15pt}}
	\label{fig:rel_ctx_write}	
\end{figure}

\begin{figure}
	\centering
	\begin{subfigure}{0.48\columnwidth}
		\includegraphics[width=\columnwidth]{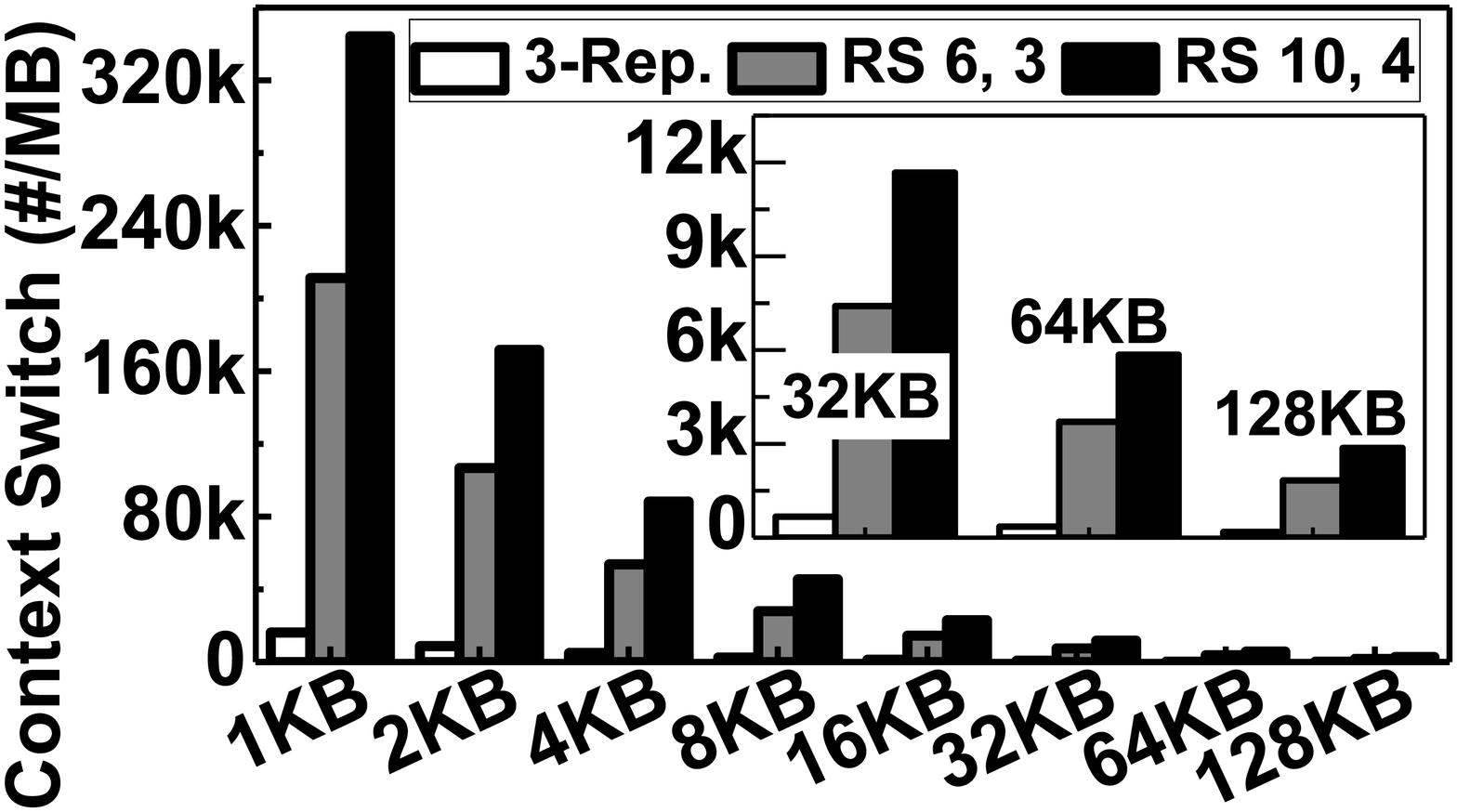}
		\caption{Sequential read.}
		\label{fig:rel_ctx_seq_read}
	\end{subfigure}
	~
	\begin{subfigure}{0.48\columnwidth}
		\includegraphics[width=\columnwidth]{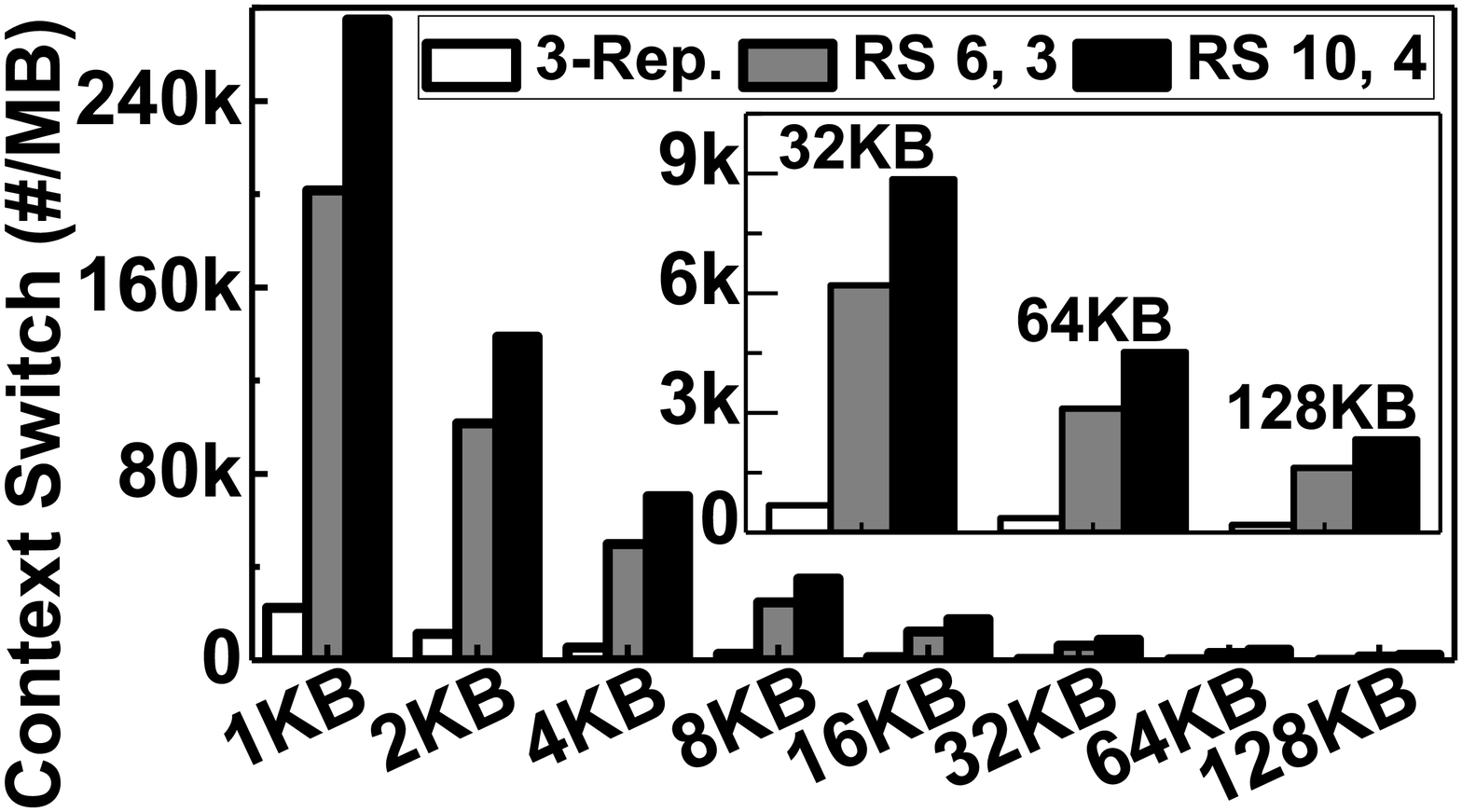}
		\caption{Random read.}
		\label{fig:rel_ctx_rand_read}
	\end{subfigure}
	\vspace{-7pt}
	\caption{relative number of context switches (reads). \vspace{-20pt}}
	\label{fig:rel_ctx_read}	
\end{figure}

\vspace{-7pt}
\subsection{Kernel and Context Switching}
\vspace{-7pt}
One of the challenges in deploying online erasure coding in distributed storage systems is user-level implementations. As we described in the previous section, user-mode operations take about 72\% of total CPU cycles for reads and writes on average. In this section, we examine context switching overheads associated with interleaving worker threads and context switches between user and kernel modes. Figures \ref{fig:rel_ctx_read} and \ref{fig:rel_ctx_write} illustrate 
relative number of context switches incurred by reads and writes. 
Since we observe that context switches occur more than 700M times per node and the number of requests processed varies due to different storage cluster performance, we use a different metric, 
relative number of context switches that is the number of context switches imposed by the storage cluster divided by the total amount of data processed. 

\vspace{-5pt}
As shown in the Figure \ref{fig:rel_ctx_write}, RS(6,3) and RS(10,4) give 4.7$\times$ and 7.1$\times$ more relative number of context switches than 3-replication for writes, respectively, on average.
We believe that this is because of two reasons. First, there is a significant amount of computing for encoding per object at the initial phase. Since Ceph manages the storage cluster with objects, even though libRBD offers block interfaces, the erasure coding module at the PG backend creates dummy data chunks and coding chunks at the initial phase and writes them to the underlying SSDs. Second, the small size of writes (smaller than the stripe width) is treated as updates. This requires reading the underlying data chunks, regenerating coding chunks and updating the corresponding stripe. This in turn introduces many activities at the cluster messenger and PG backend, which reside at the user level, thereby increasing the relative number of context switches. Note that 
RS(6,3) and RS(10,4) offer fewer 73.4\% and 73.7\% relative number of context switches for reads than for writes, respectively, but still 10$\sim$15$\times$ more than 3-replication, respectively, on average.  This is mostly because RS-concatenation related data transfers and computations, which is also performed at the user level. We will analyze why there are many context switches in online erasure coding with I/O amplification analysis and network traffic study in depth in Sections \ref{ioamplification} and \ref{private_network}.

%% file: dataoverhead.tex
\subsection{I/O Amplification}
\vspace{-8pt}
\label{ioamplification}
\begin{figure}
	\centering
	\begin{subfigure}{0.48\columnwidth}
		\includegraphics[width=\columnwidth]{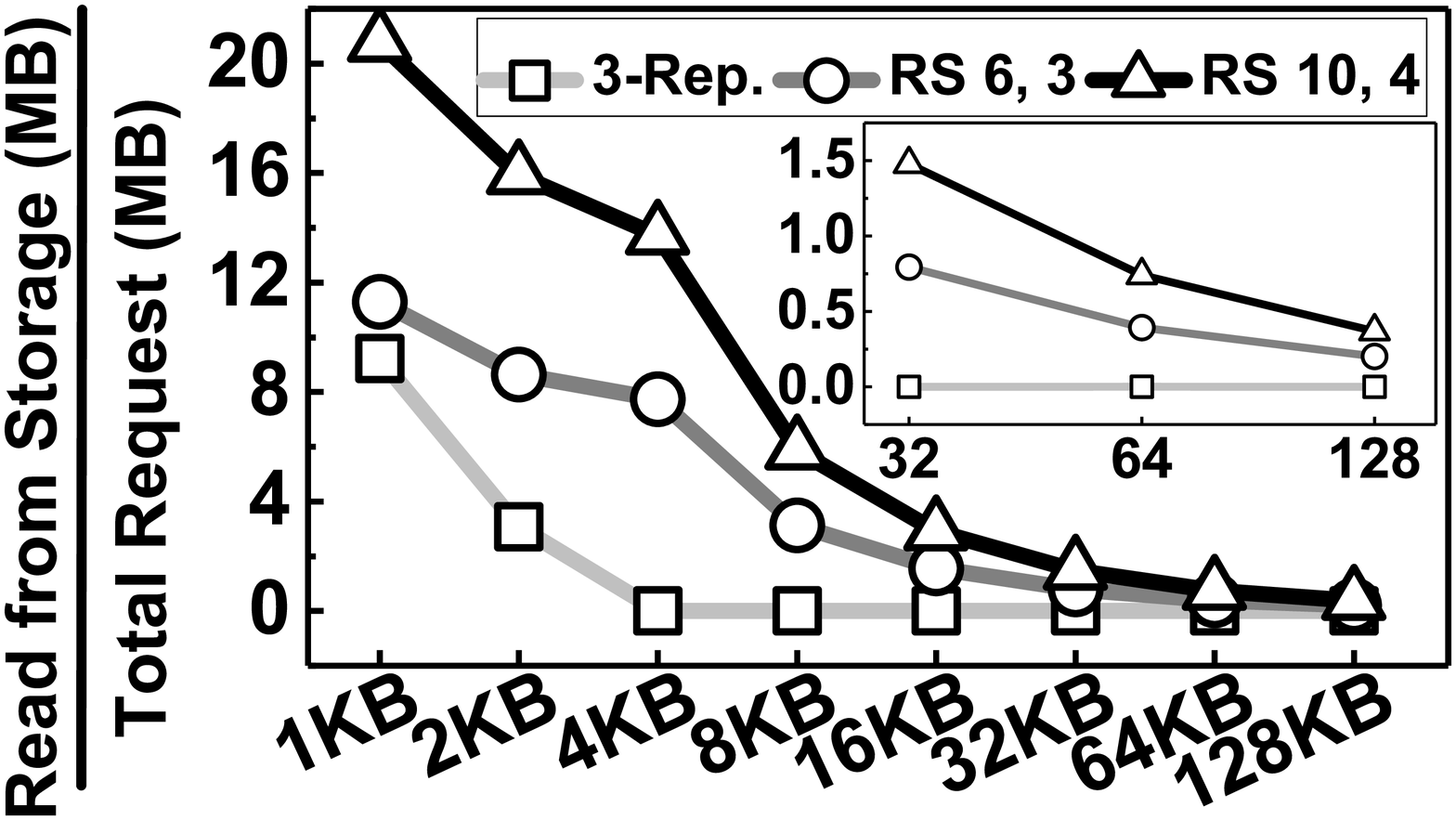}
		\caption{Reads.}
		\label{fig:seq_write_storage_oh_read}
	\end{subfigure}
	~
	\begin{subfigure}{0.48\columnwidth}	
		\includegraphics[width=\columnwidth]{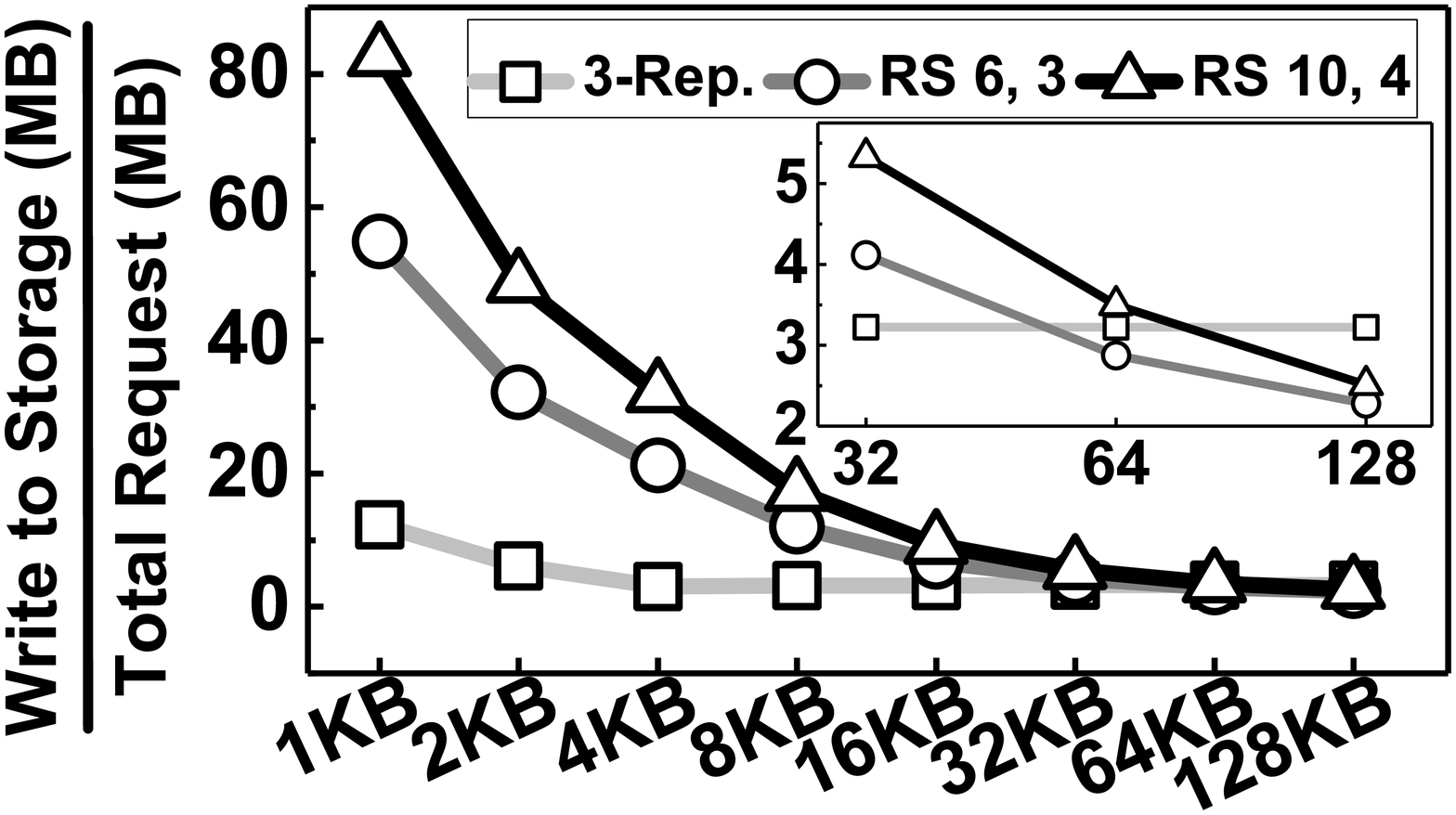}
		\caption{Writes.}
		\label{fig:seq_write_storage_oh_write}
	\end{subfigure}
	\vspace{-7pt}
	\caption{Actual storage volumes (normalized to the amount of input data) observed by sequential writes. \vspace{-12pt}}
	\label{fig:seq_write_storage_oh}
\end{figure}

\begin{figure}
	\centering
	\begin{subfigure}{0.48\columnwidth}
		\includegraphics[width=\columnwidth]{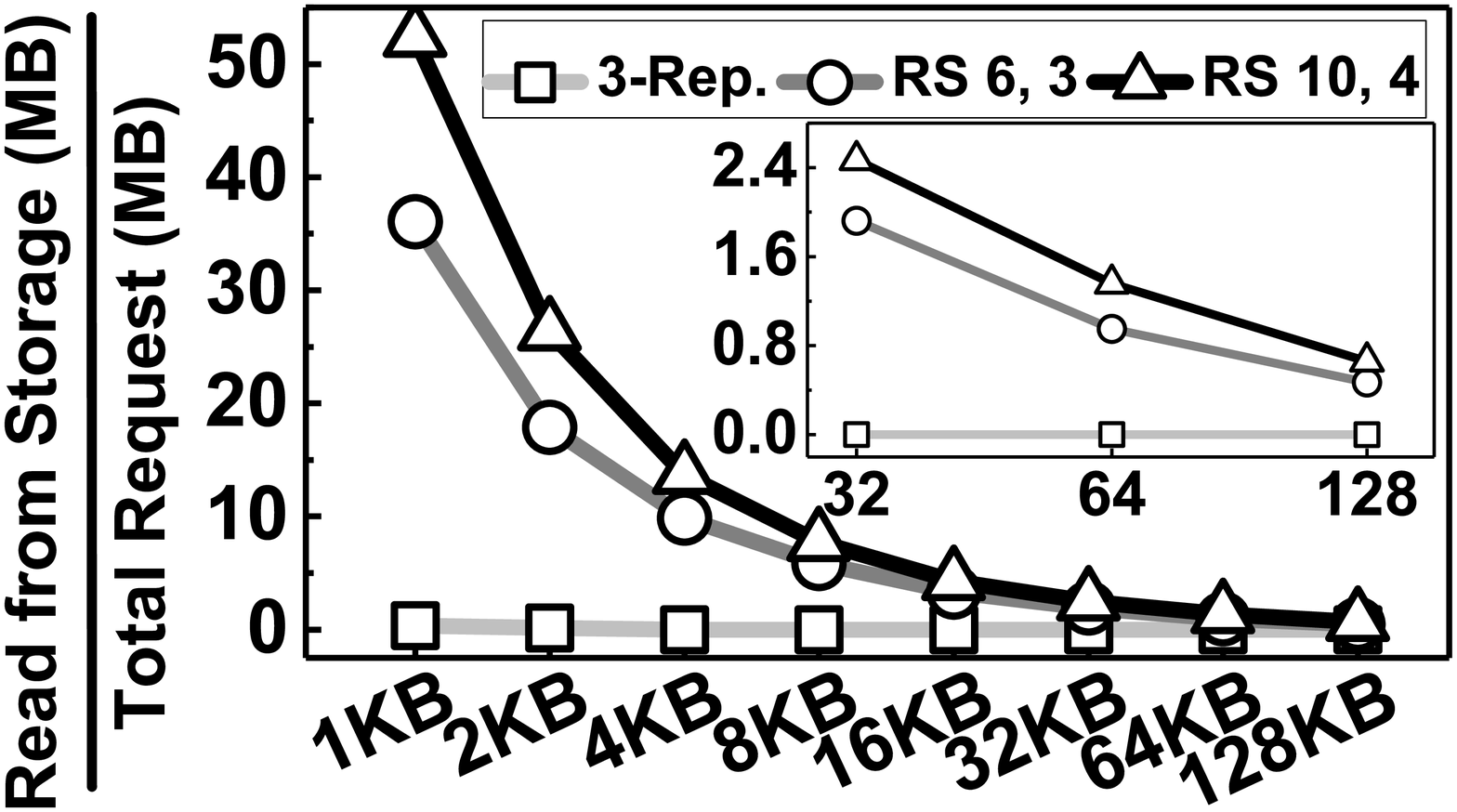}
		\caption{Reads.}
		\label{fig:rand_write_storage_oh_read}
	\end{subfigure}
	~
	\begin{subfigure}{0.48\columnwidth}
		\includegraphics[width=\columnwidth]{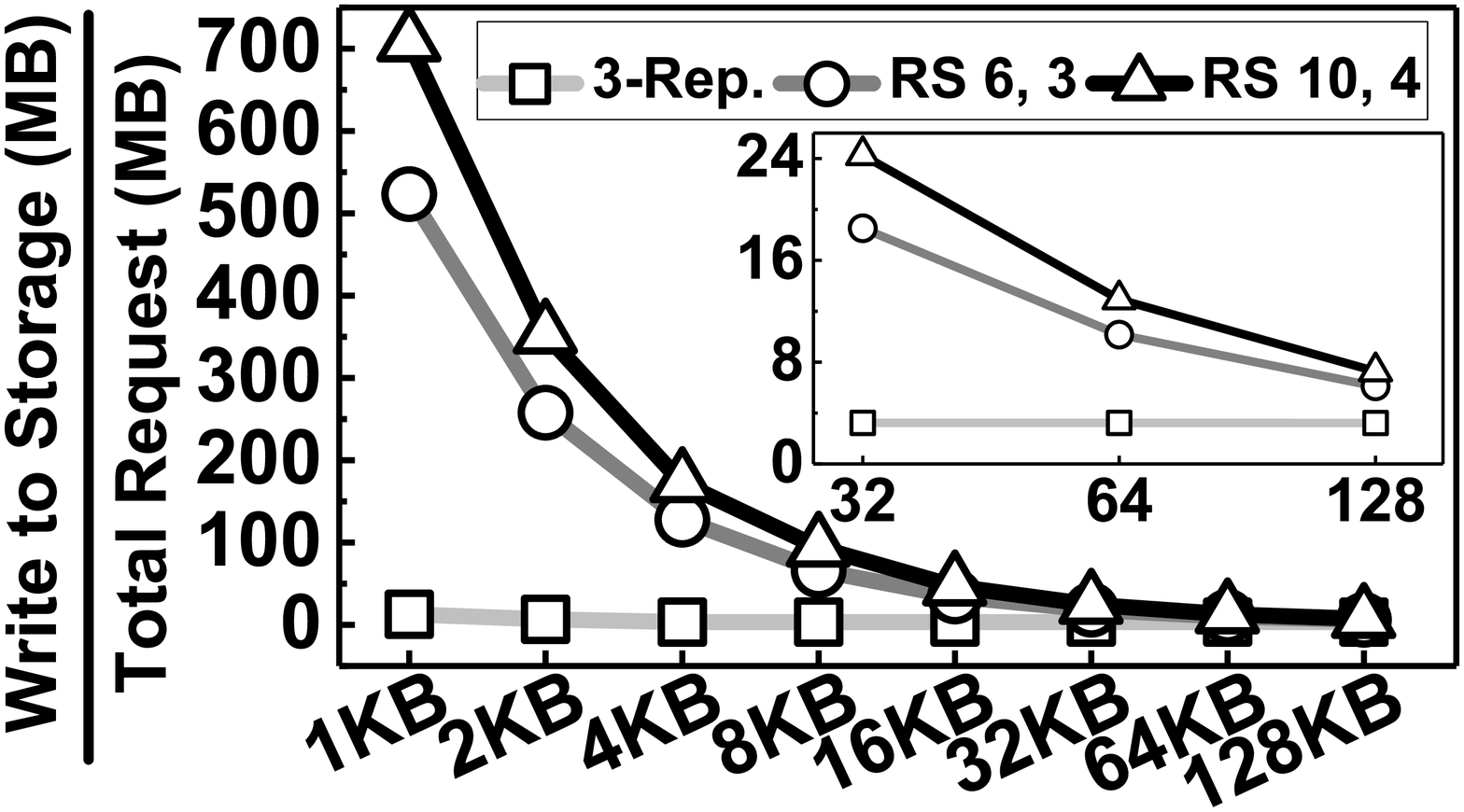}
		\caption{Writes.}
		\label{fig:rand_write_storage_oh_write}
	\end{subfigure}
	\vspace{-7pt}
	\caption{Actual storage volumes (normalized to the amount of input data) observed by random writes. \vspace{-15pt}}
	\label{fig:rand_write_storage_oh}
\end{figure}

\begin{figure}
	\centering
	\begin{subfigure}{0.48\columnwidth}
		\includegraphics[width=\columnwidth]{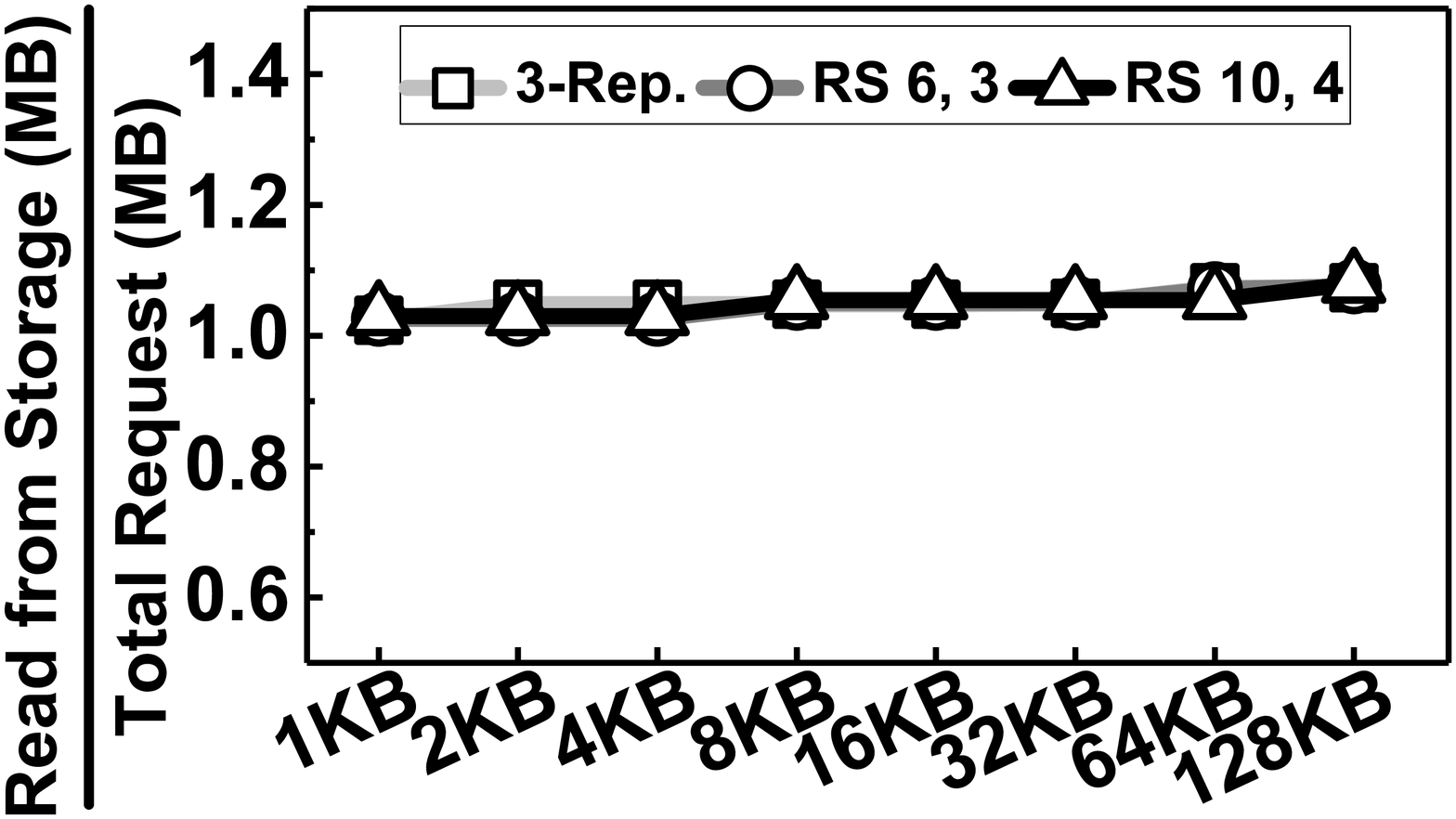}
		\caption{Sequential read.}
		\label{fig:seq_read_storage_oh}
	\end{subfigure}
	~
	\begin{subfigure}{0.48\columnwidth}
		\includegraphics[width=\columnwidth]{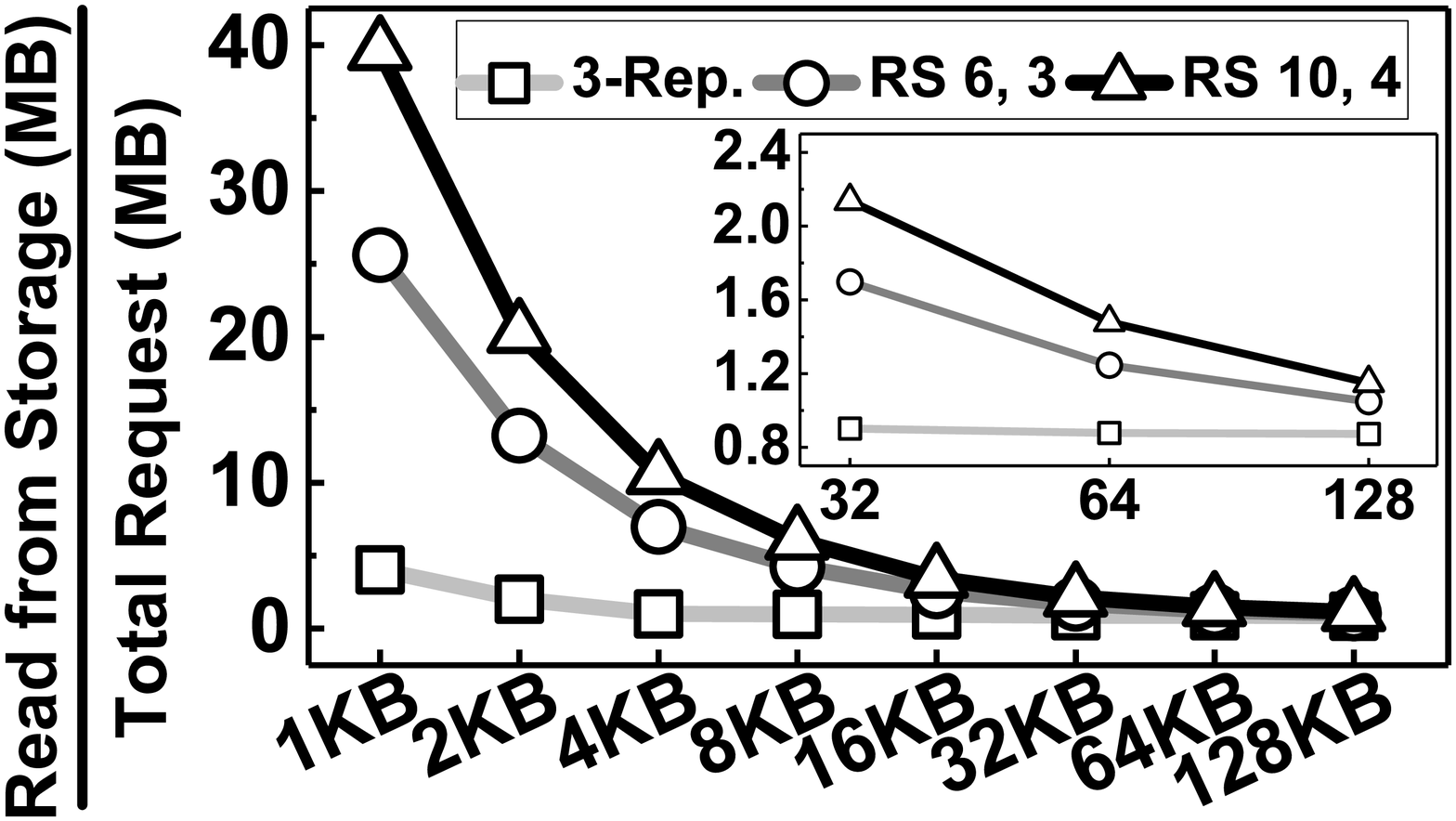}
		\caption{Random read.}
		\label{fig:rand_read_storage_oh}
	\end{subfigure}
	\vspace{-10pt}
	\caption{Actual read volumes normalized to input data. \vspace{-19pt}}
	\label{fig:read_storage_oh}
\end{figure}

\noindent \textbf{Replication vs. erasure coding.} Figures \ref{fig:seq_write_storage_oh}, \ref{fig:rand_write_storage_oh} and \ref{fig:read_storage_oh} respectively show the read and write I/O amplification calculated by normalizing the amount of data that 3-replication and erasure coding at PG backend generate with the actual amount of data that application requests. To share a high-level insight on the difference between the 3-replication and erasure coding, we first select the overheads observed by I/O services with sequential accesses; other ones will be analyzed shortly. The 3-replication generally does not require reading data when it writes, but it will perform a read-and-modify operation when the data is smaller than 4KB which is the minimum unit of I/O.
This in turn introduces read amplification 9$\times$ more than that of original request size under sequential write with 1KB block. We also observe that the I/O amplification by 3-replication is same with the amount of replicas under the writes with the blocks which size is bigger than 4KB (Figure \ref{fig:seq_write_storage_oh_write}, \ref{fig:rand_write_storage_oh_write}). 
In contrast, RS(6,3) and RS(10,4) introduce severe I/O amplification. This is because, as described in Section \ref{sec:background}, each OSD handles block requests managed by libRBD per object. Thus, even though writes occur as a sequential order (but their block size is less then the stripe), PG backend requires reading the data chunks for each write and update them with new coding chunks. As shown in the plots, this increases the amount data to read and write by up to 20.8$\times$ and 82.5$\times$ compared with total amount of requests.

\vspace{-3pt}
\noindent \textbf{Writes.} Figures \ref{fig:rand_write_storage_oh_read} and \ref{fig:rand_write_storage_oh_write} respectively show the I/O amplification under the execution of client-level random writes. As we observed from the previous subsection, RS(6,3) and RS(10,4) introduce the extra data to write more than the 3-replication up to 10.4$\times$ and many reads while replication does not need. 
As pointed out earlier, the random accesses lead to up to 55$\times$ more write amplification than 3-replication, which is completely different from the common expectation 
erasure coding on the storage overheads. Even though users write requests with small sizes, OSD requires to create an object. When users request data as random accesses, such small size requests are distributed across many OSDs and create/initialize the corresponding objects, which in turn leads to greater write amplification. We will examine this at the private network analysis (cf. Section \ref{private_network}) in detail.

\vspace{-3pt}
\noindent \textbf{Reads.} Figures \ref{fig:seq_read_storage_oh} and \ref{fig:rand_read_storage_oh} show reads with sequential and random accesses, respectively. In sequential reads, since there is no failure in the storage cluster, the PG backend for all fault tolerance mechanisms reads data chunks exactly as much as the client requests, which in turn makes the I/O amplification for all varying block sizes almost once. Even though the block size of a request is smaller than the minimum unit of I/O, consecutive I/O requests leverage the data  
and therefore there is no read amplification. However, read requests with a random pattern can be distributed and served by different OSDs. As erasure coding at the PG backend performs I/O service based on a stripe width, if the request is smaller than the stripe width, it wastes the data transfers. If a small block size request spans two stripes, the more amplification can occur. As shown in Figure \ref{fig:rand_read_storage_oh}, 
RS(6,3) and RS(10,4) impose 6.9$\times$ and 10.4$\times$ greater I/O amplification than 3-replication at 4KB. When there is a request whose block size is slightly greater than the stripe width (e.g., 32KB) and its data spans across two or three stripes, RS(6,3) and RS(10,4) impose about 2$\times$ greater I/O amplification than 3-replication.

\vspace{-10pt}
\subsection{Private Network Traffic}
\begin{figure}
	\centering
	\begin{subfigure}[b]{0.48\columnwidth}
		\includegraphics[width=\columnwidth]{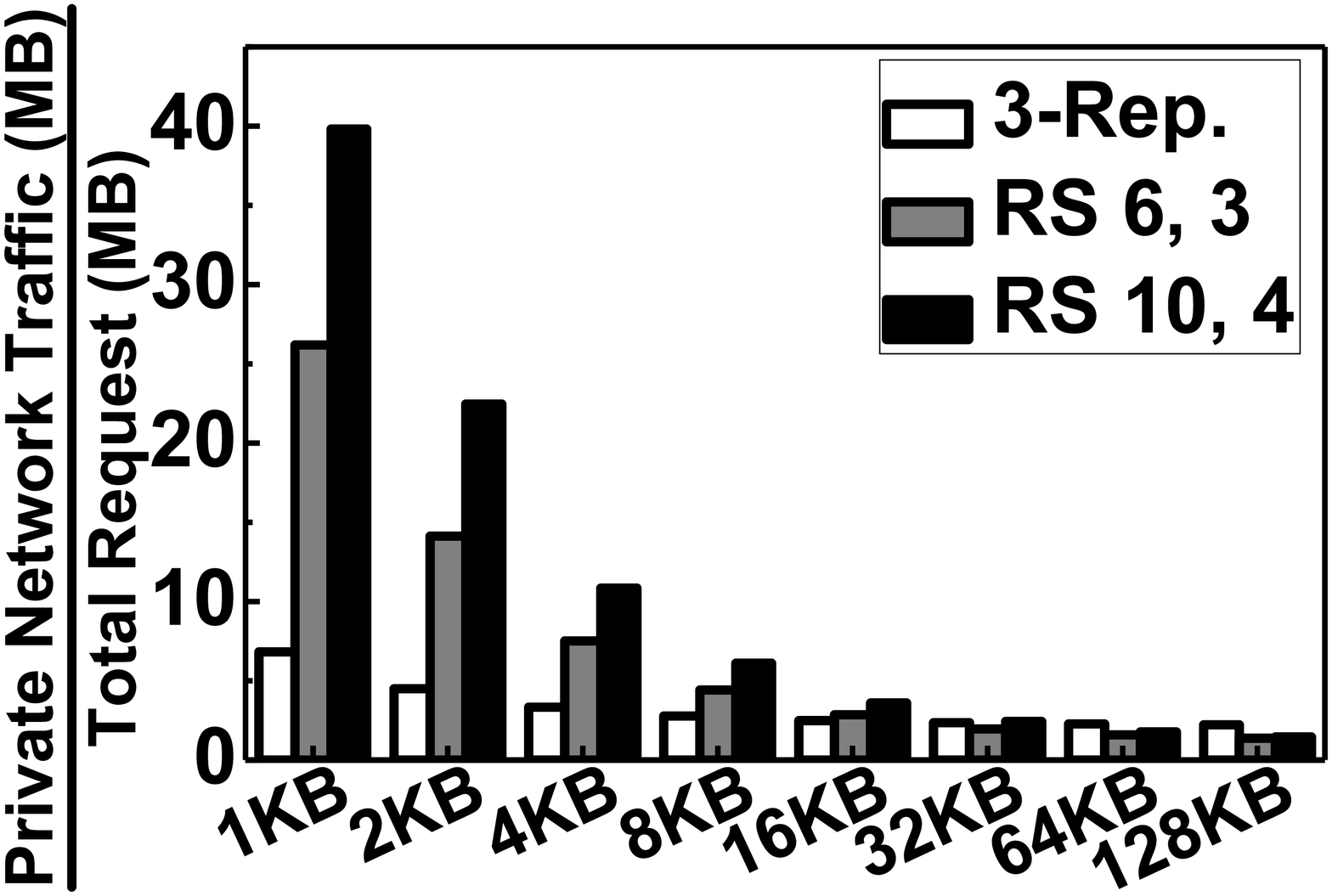}
		\caption{Sequential write.}
		\label{fig:priv_net_seq_write}
	\end{subfigure}
	~
	\begin{subfigure}[b]{0.48\columnwidth}
		\includegraphics[width=\columnwidth]{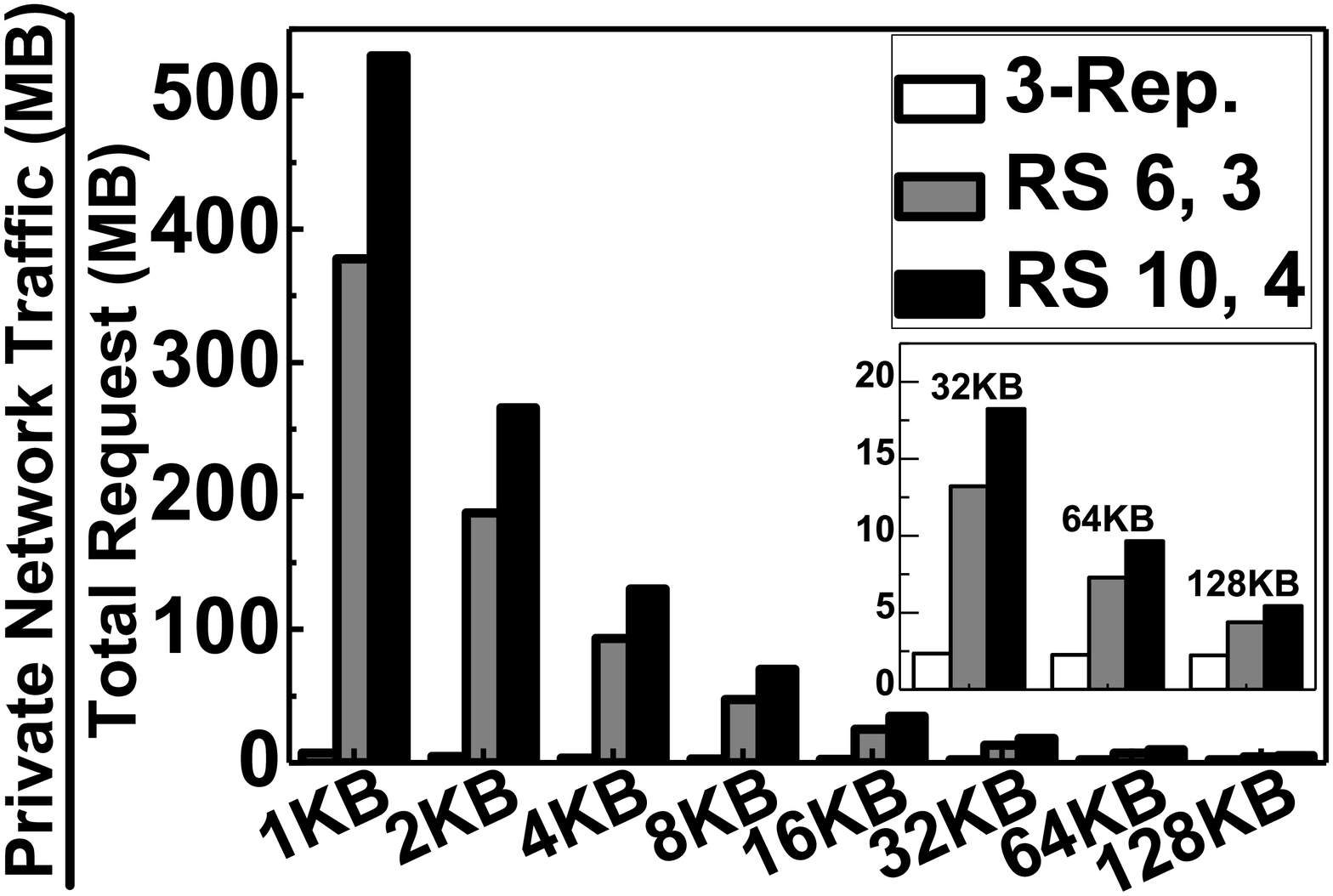}
		\caption{Random write.}
		\label{fig:priv_net_rand_write}
	\end{subfigure}
	\vspace{-20pt}
	\caption{Private network overhead analysis for writes. \vspace{-15pt}}
	\label{fig:priv_net_write}
\end{figure}

\vspace{-7pt}
\label{private_network}
\noindent \textbf{Writes.} Ones may expect that 3-replication exhibits more data transfers as the amount of replica data is more than coding chunks of RS(6,3) and RS(10,4). However, 
under writes with various block sizes and patterns, the I/O amplification imposed by RS(6,3) and RS(10,4) make the private network congested. Specifically, as shown in Figure \ref{fig:priv_net_write}, RS(6,3) and RS(10,4) generate 2.4$\times$ and 3.5$\times$ more data transfers over the actual request size than 3-replication, respectively, if the block size is smaller than 32KB. 
As the block size increases, the portion of extra stripes decreases, and it results in the decrease of private network traffic. Importantly, under the random accesses, the private network is overly congested by RS(6,3) and RS(10,4). As shown in Figure \ref{fig:priv_net_rand_write}, 
RS(6,3) and RS(10,4) transfer 53.3$\times$ and 74.7$\times$ more data than 3-replication, which should be optimized for a future system. This is because erasure coding of the PG backend requires initializing the objects, which can be significant overheads when the requests are distributed across multiple OSDs. 

\begin{figure}
	\centering
	\begin{subfigure}[b]{0.48\columnwidth}
		\includegraphics[width=\columnwidth]{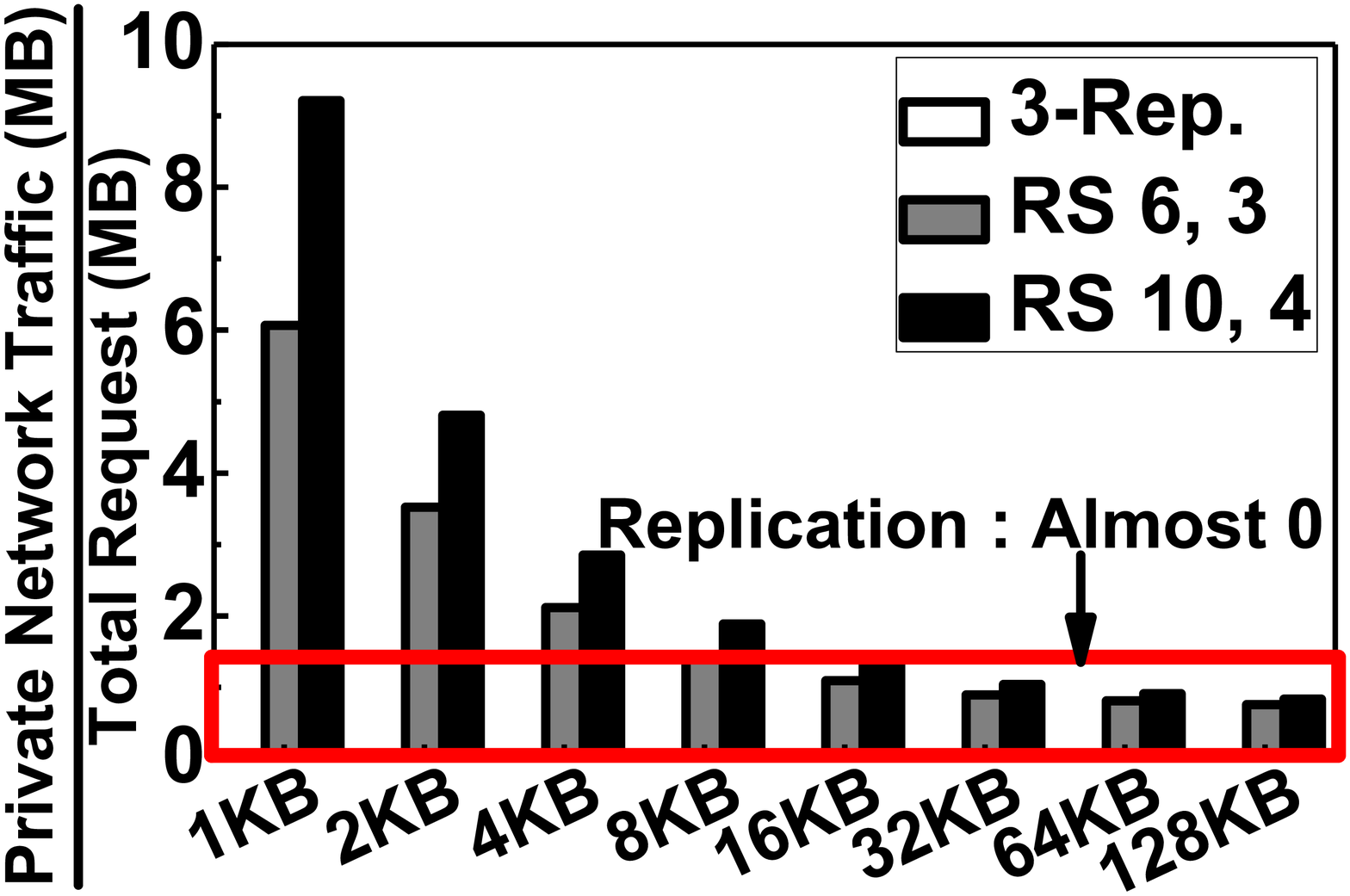}
		\caption{Sequential read.}
		\label{fig:priv_net_seq_read}
	\end{subfigure}
	~
	\begin{subfigure}[b]{0.48\columnwidth}
		\includegraphics[width=\columnwidth]{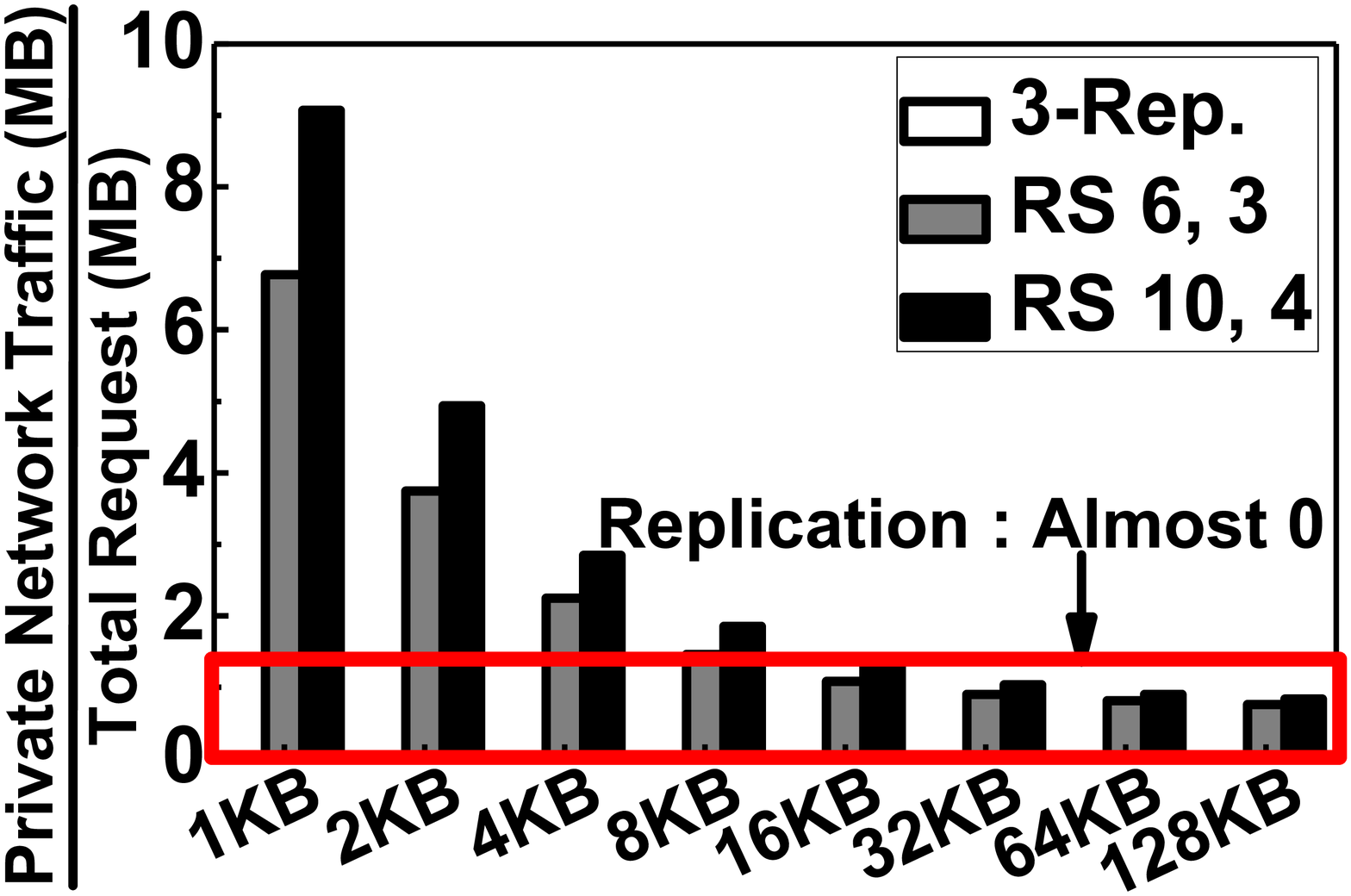}
		\caption{Random read.}
		\label{fig:priv_net_rand_read}
	\end{subfigure}
	\vspace{-20pt}
	\caption{Private network overhead analysis for reads. \label{fig:priv_net_read} \vspace{-15pt}}
\end{figure}

\noindent \textbf{Reads.} Figure \ref{fig:priv_net_read} shows the data transfer overheads imposed by the 
fault tolerance mechanisms of the PG backend during read services. 
From this plot, we can observe that 3-replication exhibits only minimum data transfers related to necessary communications among OSD daemons. This OSD interaction is used for monitoring the status of each OSD, referred to as OSD heartbeat. It generates 20KB/s data transfers in this work, which has almost no impact for the private network traffic. In contrast, RS(6,3) and RS(10,4) introduce data transfers upto
6.8$\times$ and 9.1$\times$ more than that of original request size which is associated with concatenating data chunks into a stripe. 
Note that the amount of transferred data (for both reads and writes) over the network cannot be statistically calculated by considering the relationship between $k$ data chunks and $m$ coding chunks since there is often a case that the data is transferred among OSDs within a node.

%% file: distribution.tex
In this section, we will examine the performance impacts of erasure coding imposed by different data distributions and object managements of underlying parallel file systems. 

\vspace{-8pt}
\subsection{Placement Group Parallelism}
\begin{figure}
	\centering
	\begin{subfigure}{0.48\columnwidth}
		\includegraphics[width=\columnwidth]{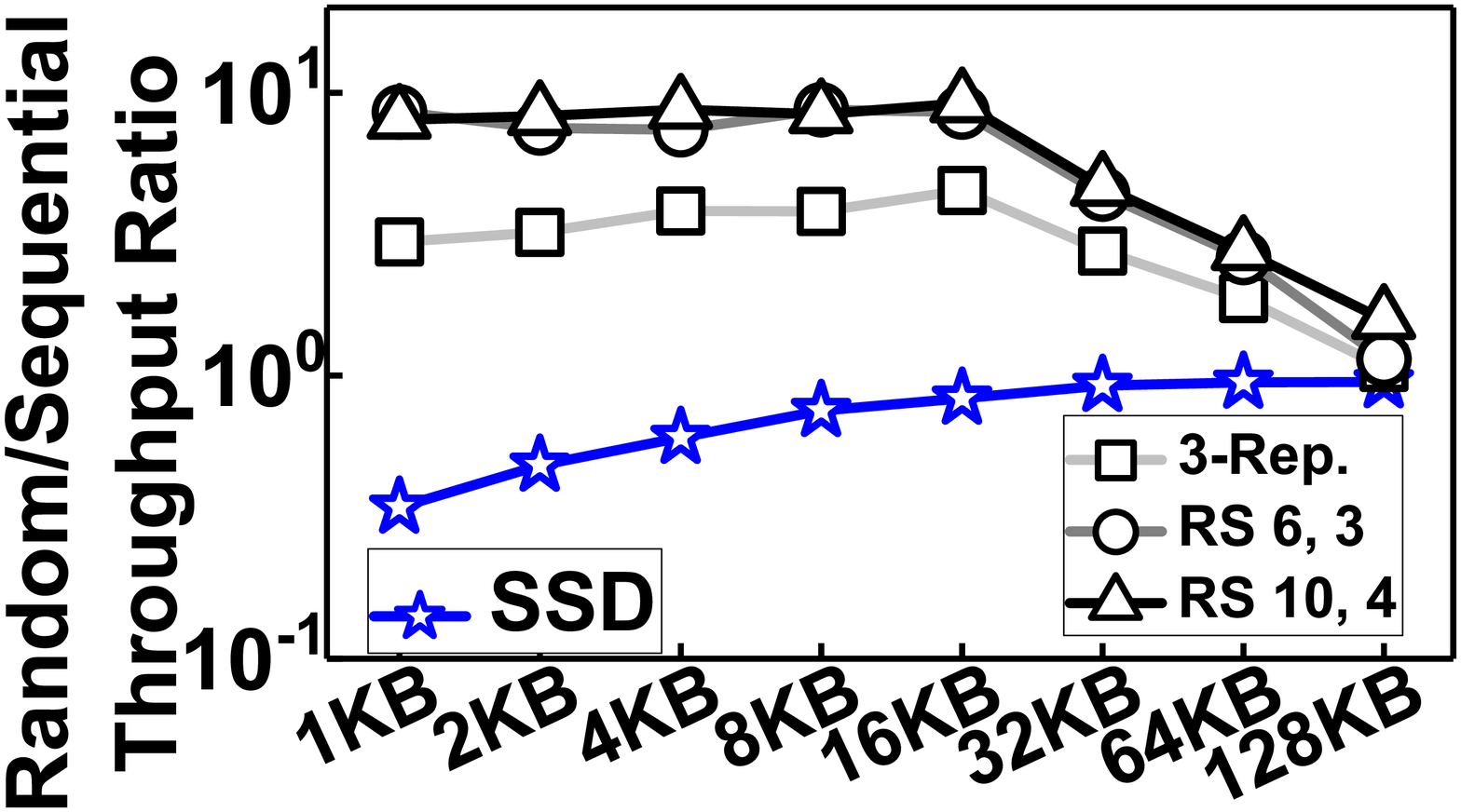}
		\caption{Read.}
		\label{fig:parallelism_read}
	\end{subfigure}
	~
	\begin{subfigure}{0.48\columnwidth}
		\includegraphics[width=\columnwidth]{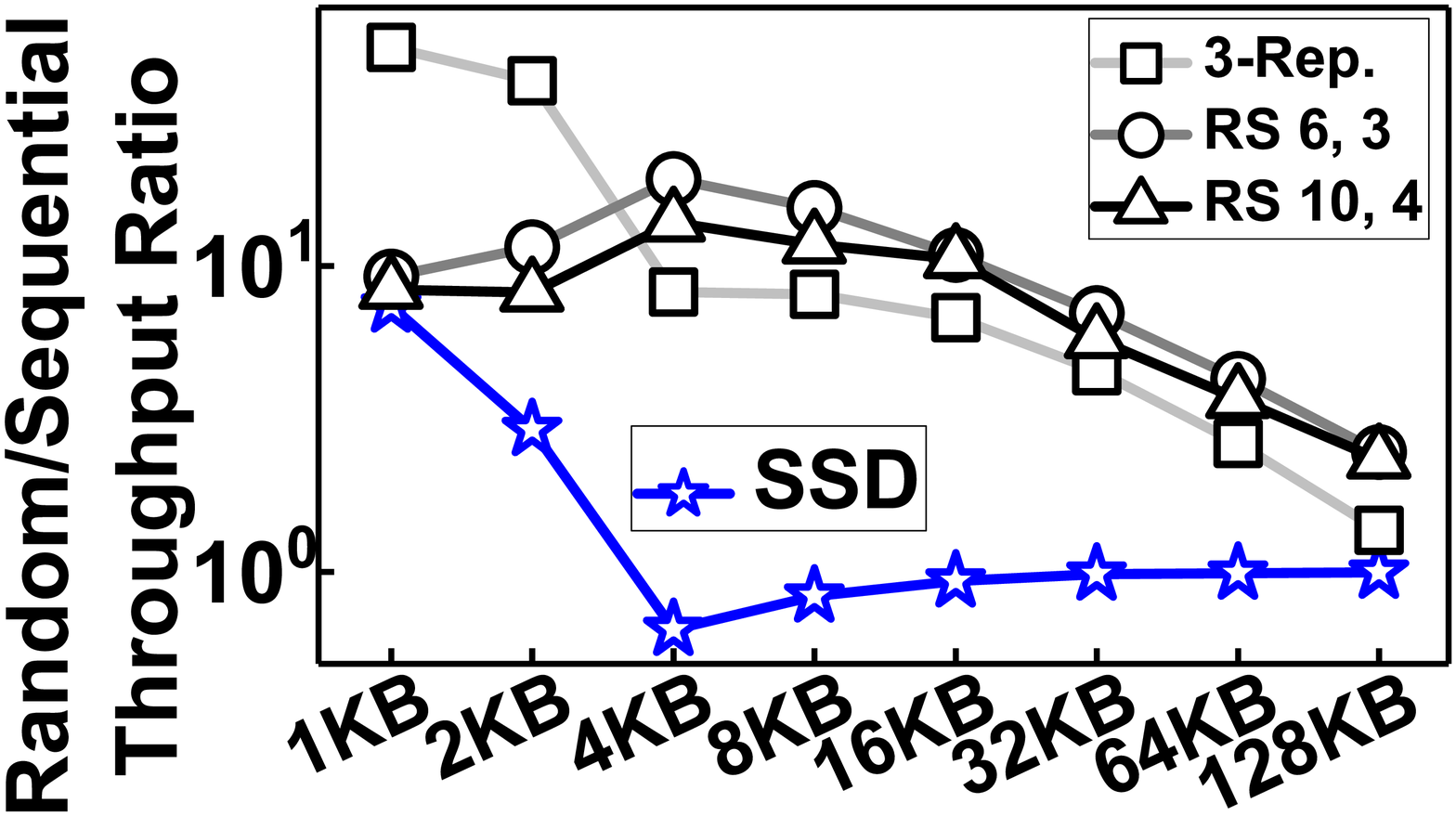}
		\caption{Write.}
		\label{fig:parallelism_write}
	\end{subfigure}
	\vspace{-7pt}
	\caption{Random/sequential throughput ratios. \vspace{-12pt}}
	\label{fig:rand_seq_ratio}
\end{figure}

\vspace{-8pt}
Figure \ref{fig:rand_seq_ratio} illustrates the ratio of throughput observed under random accesses to that under sequential accesses (i.e., random/sequential ratio). In this evaluation, we evaluate a bare SSD (used for the OSD composition) without any fault tolerance mechanism and compare random/sequential ratios of the SSD with those of the underlying storage cluster with 3-replication, RS(6,3) and RS(10,4). From this plot, we can observe that random/sequential ratios of the SSD are mostly close to one for the block sizes greater than 4KB, and they are actually negative for the 1$\sim$4KB request sizes. These results mean the bare SSD's throughput under random accesses is similar to or worse than that of sequential accesses. However, the random/sequential ratios of the storage cluster behave completely different story compared with the bare SSD. 
For small size of requests, the storage cluster which employs 3-replication exhibits bigger random/sequential throughput ratio then the bare SSD by 7 times, on average.  Moreover, the storage clusters with RS(6,3) and RS(10,4) offer even 2.3 times and 2.5 times better than the cluster with 3-replication, on average. We believe that this is because erasure codes of PG backend distribute the data across different OSDs, which introduce less contentions on a primary OSD and increase the parallelism. For the writes (Figure \ref{fig:parallelism_write}), we can see the same behaviors we observed in read performance analysis. Specifically, RS(6,3) and RS(10,4) show 3.7$\times$ and 2.8$\times$ higher random throughput than the bare SSD, respectively, on average.

\begin{figure}
	\centering
	\includegraphics[width=\columnwidth]{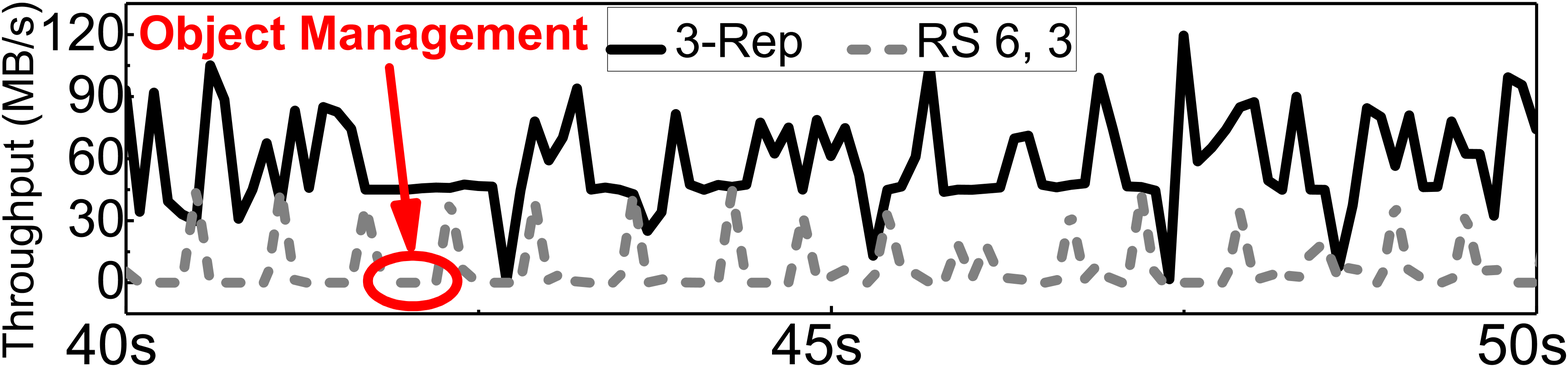}
	\vspace{-18pt}
	\caption{Time series analysis for sequential write.\vspace{-17pt}}
	\label{fig:pub_bw_time}
\end{figure}

\vspace{-10pt}
\subsection{Object Management}
\vspace{-7pt}

Since the storage cluster manages data over an object, it gives a burden to erasure coding at some extent. Specifically, while libRBD offers block interface operations, the data is managed by storage nodes as an object. Thus, whenever an OSD daemon receives a new write that heads a target address in the range of an object, it creates the object and fills the data shards and coding shards. This object initialization degrades the overall system performance. Figure \ref{fig:pub_bw_time} shows a time series analysis for the initial performance by issuing 16KB data as a sequential pattern. One can observe from the figure that, while 3-replication doesn't have performance degradation at the I/O initial phase, RS(6,3) periodically shows near-zero throughput due to the object initialization that fills data and coding shards. 

\begin{figure}
	\centering
	\includegraphics[width=\columnwidth]{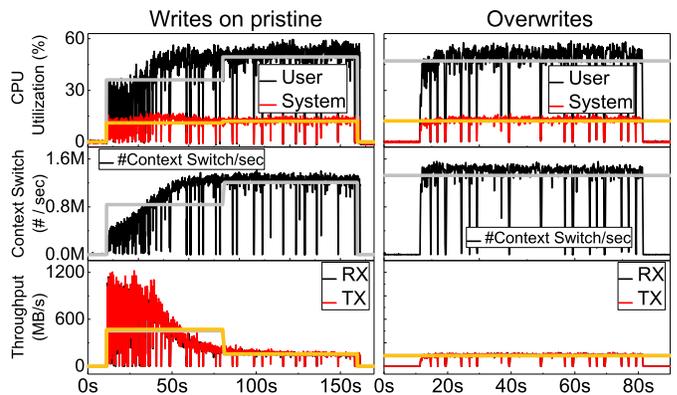}
	\caption{Time series analysis for CPU utilization, the number of context switches per second, and private network throughput observed by random writes. \vspace{-18pt}}
	\label{fig:pristine_over_compare}
\end{figure}

To be precise, we further analyze the performance overheads imposed by initializing data and coding shards. Figure \ref{fig:pristine_over_compare} shows the time series analysis by comparing writes on a pristine image (left) and overwrites (right) in terms of the CPU utilizations, context switch overheads, and private network traffic. This time series analysis captures the values from the beginning of I/O process to the point where the performance of storage cluster gets stabilized. 
From the plot, we can observe that the CPU utilizations and context switches observed by random writes on the pristine image are lower than those of overwrites by 20\%, 37\%, respectively, on average until 70 sec, and they converges after 70 sec. At the same time period, the private network (of writes on the clean image) is busy to transfer/receive data from each other until 70 sec, which is 3.5$\times$ greater than that of overwrites on average. The data on overwrites were already written in the target, and therefore, there is no data transfer for object initialization, which in turn makes the CPU utilization and context switch high at the very beginning of the I/O processes.

%% file: discussion.tex
Replication is the most common but effective and practical method to maintain the storage reliability and resilience \cite{tanenbaum2007distributed}. Since the storage overheads imposed by replicas make distributed storage systems non-scalable and the network overly crowded, many studies pay attention on erasure coding for distributed file systems \cite{rashmi2013solution, rashmi2016ec, mitra2016partial, silberstein2014lazy}. However, erasure codes also suffer from heavy overheads in cases where they need to reconstruct data on a node or device failure. This is a well known problem, called repair bandwidth, which not only significantly degrades system performance \cite{rashmi2013solution} but also severely amplifies data volume and network traffic. To address these challenges, new techniques on erasure codes are propsed \cite{rashmi2015having, rashmi2016ec, mitra2016partial}. EC-Cache \cite{rashmi2016ec} is designed to maximize the load balancing and reduce the latency for erasure codes. \cite{silberstein2014lazy} addressed the repair bandwidth issue with a lazy recovery mechanism. While all these studies analyzed the overheads on data reconstruction and optimized decoding procedures of erasure coding applied to disk-based storage clusters, our work highlighted many observations with in-depth study for encoding mechanisms on distributed storage systems and parallel file system implementation.

\vspace{-3pt}
On the other hand, \cite{mohan2015benchmarking} reports that the performance of erasure coding on Hadoop distributed file system (HDFS) does not degrade, compared to that of replications.
This study was performed on a HDD-based cluster, and HDFS applies I/O striping to both erasure coding and replication. In contrast, Ceph used in this work is optimized for a SSD-based cluster, which directly service I/Os from a target OSD. However, erasure coding has software interventions due to its en/decoding related computations and raises synchronization/barrier issues, which makes it employ I/O striping rather than the direct I/O services. This introduces further performance differences between erasure coding and replication on SSDs, which is not observed by previous studies (on HDDs). 

\noindent \textbf{Limits of this study.} We believe that applying erasure coding to SSD arrays is in an early stage for parallel file systems, and some observations of this work may depend on various implementations of them. However, several fundamental behaviors that erasure coding requires such as matrix multiplications with new and old data will introduce similar challenges that we reported in this paper. For example, erasure coding implemented at user-level needs to read data and generates new parity bits, which introduces not only extra I/O activities such as network service and storage I/O requests but also system overheads such as kernel mode switching and memory copies. In addition, if the target file system changes its configuration such as a stripe width, actual values on our observations would also change, but their trend will not vary too much. For example, in cases where the target system increases a stripe width, latencies of both encoding and decoding almost linearly increase, which is also observed by other studies \cite{ hu2012nccloud, papailiopoulos2012simple}. 



%% file: acknowledgement.tex
This research is mainly supported by NRF 2016R1C1B2015312. This work is also supported in part by IITP-2017-2017-0-01015, NRF-2015M3C4A7065645, DOE DE-AC02-05CH 11231, and MemRay grant (2015-11-1731). Nam Sung Kim is supported in part by NSF 1640196 and SRC/NRC NERC 2016-NE-2697-A. Myoungsoo Jung is the corresponding author.

%% file: conclusion.tex
In this work, we studied the overheads imposed by erasure coding on a distributed SSD array system. In contrast to the common expectation on erasure codes, we observed that they exhibit heavy network traffic (which is invisible to users) greater than a popular replication method (triple replicas) by 75 times at most, which in turn also increase the amount of actual data that the underlying SSD needs to manage. This work also reveals that the erasure coding mechanisms on the distributed SSD array systems introduce 21 times more context switches and require 12 times more CPU cycles at most, than the replication due to user-level implementation and storage cluster management. 

%% file: main.bbl
\begin{thebibliography}{10}
\providecommand{\url}[1]{#1}
\csname url@samestyle\endcsname
\providecommand{\newblock}{\relax}
\providecommand{\bibinfo}[2]{#2}
\providecommand{\BIBentrySTDinterwordspacing}{\spaceskip=0pt\relax}
\providecommand{\BIBentryALTinterwordstretchfactor}{4}
\providecommand{\BIBentryALTinterwordspacing}{\spaceskip=\fontdimen2\font plus
\BIBentryALTinterwordstretchfactor\fontdimen3\font minus
  \fontdimen4\font\relax}
\providecommand{\BIBforeignlanguage}[2]{{%
\expandafter\ifx\csname l@#1\endcsname\relax
\typeout{** WARNING: IEEEtranS.bst: No hyphenation pattern has been}%
\typeout{** loaded for the language `#1'. Using the pattern for}%
\typeout{** the default language instead.}%
\else
\language=\csname l@#1\endcsname
\fi
#2}}
\providecommand{\BIBdecl}{\relax}
\BIBdecl

\bibitem{cephpoolconfig}
``Ceph documentation,'' \emph{Pool, PG and CRUSH config referenceâ€, Ceph,
  URL:
  http://docs.ceph.com/docs/master/rados/configuration/pool-pg-config-ref/}.

\bibitem{cephkrakenrelease}
``Ceph homepage,'' \emph{Ceph 11.2.0-"Kraken Release"â€, Ceph, URL:
  http://ceph.com/releases/v11-2-0-kraken-released/}, 2017.

\bibitem{axboe2015flexible}
J.~Axboe, ``Flexible io tester,'' 2015.

\bibitem{borthakur2010hdfs}
D.~Borthakur \emph{et~al.}, ``Hdfs raid,'' in \emph{Hadoop User Group Meeting},
  2010.

\bibitem{brunelle2006block}
A.~D. Brunelle, ``Block i/o layer tracing: blktrace,'' \emph{HP,
  Gelato-Cupertino, CA, USA}, 2006.

\bibitem{chen2009understanding}
F.~Chen \emph{et~al.}, ``Understanding intrinsic characteristics and system
  implications of flash memory based solid state drives,'' in
  \emph{SIGMETRICS}, 2009.

\bibitem{ref-raid}
Dell, \emph{PowerEdge RAID Controller H730P}, 2014.

\bibitem{dimakis2011survey}
A.~G. Dimakis \emph{et~al.}, ``A survey on network codes for distributed
  storage,'' \emph{IEEE}, 2011.

\bibitem{esmaili2013core}
K.~S. Esmaili \emph{et~al.}, ``Core: Cross-object redundancy for efficient data
  repair in storage systems,'' in \emph{IEEE bigdata}, 2013.

\bibitem{ford2010availability}
D.~Ford \emph{et~al.}, ``Availability in globally distributed storage
  systems.'' in \emph{OSDI}, 2010.

\bibitem{ghemawat2003google}
S.~Ghemawat \emph{et~al.}, ``The google file system,'' in \emph{ACM SIGOPS
  operating systems review}, 2003.

\bibitem{greenberg2008cost}
A.~Greenberg \emph{et~al.}, ``The cost of a cloud: research problems in data
  center networks,'' \emph{ACM SIGCOMM computer communication review}, 2008.

\bibitem{hu2009write}
X.-Y. Hu \emph{et~al.}, ``Write amplification analysis in flash-based solid
  state drives,'' in \emph{SYSTOR}, 2009.

\bibitem{hu2012nccloud}
Y.~Hu \emph{et~al.}, ``Nccloud: applying network coding for the storage repair
  in a cloud-of-clouds.'' in \emph{FAST}, 2012.

\bibitem{huang2012erasure}
C.~Huang \emph{et~al.}, ``Erasure coding in windows azure storage.'' in
  \emph{Usenix ATC}, 2012.

\bibitem{jung2013revisiting}
M.~Jung \emph{et~al.}, ``Revisiting widely-held expectations of ssd and
  rethinking implications for systems,'' \emph{SIGMETRICS}, 2013.

\bibitem{jung2016exploring}
M.~Jung, ``Exploring design challenges in getting solid state drives closer to
  cpu,'' \emph{TC}, 2016.

\bibitem{lacan2004systematic}
J.~Lacan \emph{et~al.}, ``Systematic mds erasure codes based on vandermonde
  matrices,'' \emph{IEEE Communications Letters}, 2004.

\bibitem{metz2012google}
C.~Metz, ``Google remakes online empire with â€˜colossusâ€™,'' \emph{Wired
  [Online]. Available: http://www. wired.
  com/2012/07/google-colossus/[Accessed: May 4, 2014]}, 2012.

\bibitem{mitra2016partial}
S.~Mitra \emph{et~al.}, ``Partial-parallel-repair (ppr): a distributed
  technique for repairing erasure coded storage,'' in \emph{EuroSys}, 2016.

\bibitem{mohan2015benchmarking}
L.~J. Mohan \emph{et~al.}, ``Benchmarking the performance of hadoop triple
  replication and erasure coding on a nation-wide distributed cloud,'' in
  \emph{NetCod}, 2015.

\bibitem{omura1986computational}
J.~K. Omura \emph{et~al.}, ``Computational method and apparatus for finite
  field arithmetic,'' May~6 1986, uS Patent 4,587,627.

\bibitem{papailiopoulos2012simple}
D.~S. Papailiopoulos \emph{et~al.}, ``Simple regenerating codes: Network coding
  for cloud storage,'' in \emph{INFOCOM}, 2012.

\bibitem{plank2014jerasure}
J.~S. Plank \emph{et~al.}, ``Jerasure: A library in c facilitating erasure
  coding for storage applications--version 2.0,'' Technical Report
  UT-EECS-14-721, University of Tennessee, Tech. Rep., 2014.

\bibitem{rashmi2013solution}
K.~Rashmi \emph{et~al.}, ``A solution to the network challenges of data
  recovery in erasure-coded distributed storage systems: A study on the
  facebook warehouse cluster.'' in \emph{HotStorage}, 2013.

\bibitem{rashmi2015having}
K.~RAshmi \emph{et~al.}, ``Having your cake and eating it too: Jointly optimal
  erasure codes for i/o, storage, and network-bandwidth.'' in \emph{FAST},
  2015.

\bibitem{rashmi2016ec}
K.~RASHMI \emph{et~al.}, ``Ec-cache: load-balanced, low-latency cluster caching
  with online erasure coding,'' in \emph{OSDI}, 2016.

\bibitem{reed1960polynomial}
I.~S. Reed \emph{et~al.}, ``Polynomial codes over certain finite fields,''
  \emph{SIAM journal}, 1960.

\bibitem{rosenthal1999maximum}
J.~Rosenthal \emph{et~al.}, ``Maximum distance separable convolutional codes,''
  \emph{Applicable Algebra in Engineering, Communication and Computing}, 1999.

\bibitem{sathiamoorthy2013xoring}
M.~Sathiamoorthy \emph{et~al.}, ``Xoring elephants: Novel erasure codes for big
  data,'' in \emph{PVLDB}, 2013.

\bibitem{shahidi2016exploring}
N.~Shahidi \emph{et~al.}, ``Exploring the potentials of parallel garbage
  collection in ssds for enterprise storage systems,'' in \emph{SC}, 2016.

\bibitem{silberstein2014lazy}
M.~Silberstein \emph{et~al.}, ``Lazy means smart: Reducing repair bandwidth
  costs in erasure-coded distributed storage,'' in \emph{SYSTOR}, 2014.

\bibitem{tanenbaum2007distributed}
A.~S. Tanenbaum \emph{et~al.}, \emph{Distributed systems: principles and
  paradigms}.\hskip 1em plus 0.5em minus 0.4em\relax Prentice-Hall, 2007.

\bibitem{weil2006ceph}
S.~A. Weil \emph{et~al.}, ``Ceph: A scalable, high-performance distributed file
  system,'' in \emph{OSDI}, 2006.

\bibitem{weil2006crush}
S.~A. WEil \emph{et~al.}, ``Crush: Controlled, scalable, decentralized
  placement of replicated data,'' in \emph{SC}, 2006.

\bibitem{weil2007rados}
S.~A. WEIL \emph{et~al.}, ``Rados: a scalable, reliable storage service for
  petabyte-scale storage clusters,'' in \emph{international workshop on
  Petascale data storage}, 2007.

\bibitem{zhang2015opennvm}
J.~Zhang \emph{et~al.}, ``Opennvm: An open-sourced fpga-based nvm controller
  for low level memory characterization,'' in \emph{ICCD}, 2015.

\end{thebibliography}
